\documentclass[journal]{IEEEtran}

\usepackage{cite}
\usepackage{amsmath,amssymb,amsfonts}
\usepackage{bm}
\usepackage{mathtools}
\usepackage{array}
\usepackage{booktabs}
\usepackage{multirow}
\usepackage{makecell}
\usepackage{tabularx}
\usepackage{graphicx}
\usepackage{url}
\usepackage{xcolor}
\usepackage[caption=false,font=footnotesize]{subfig}
\usepackage{siunitx}
\DeclareSIUnit{\dBm}{dBm}
\DeclareSIUnit{\dB}{dB}
\usepackage{algorithm}
\usepackage{algorithmic}
\usepackage{stfloats}
\usepackage{balance}
\usepackage[hidelinks]{hyperref}

\DeclareMathOperator*{\argmax}{arg\,max}

\graphicspath{{./}}

\title{Device-Centric Cooperative Floor-Level Estimation under Sparse Anchor Deployment for Construction-Site Applications}

\author{
Jiaxu Liang,
Zhuoran Li,
Osamu Muta,
Haruto Morokuma,
Naohiro Ikeda,
and Takamasa Kitanouma%
\thanks{Jiaxu Liang, Zhuoran Li, and Haruto Morokuma are with the Graduate School of Information Science and Electrical Engineering, Kyushu University, Fukuoka, Japan.}
\thanks{Osamu Muta is with the Faculty of Information Science and Electrical Engineering, Kyushu University, Fukuoka, Japan.}
\thanks{Naohiro Ikeda is with the Institute of Technology, Tokyu Construction Co., Ltd., Sagamihara, Kanagawa, Japan.}
\thanks{Takamasa Kitanouma is with Phindex Technologies, Suita, Osaka, Japan.}
\thanks{Corresponding authors: Jiaxu Liang and Osamu Muta
(e-mail: j.liang@mobcom.ait.kyushu-u.ac.jp; muta@ieee.org; muta@ait.kyushu-u.ac.jp).}
\thanks{This work has been submitted to the IEEE for possible publication.
Copyright may be transferred without notice, after which this version may no longer be accessible.}

}

\begin{document}
\maketitle

\begin{abstract}
For construction-site applications in multi-story buildings, accurate floor-level estimation is important for tracking workers, equipment, and Internet of Things (IoT) devices. In buildings under construction, communication infrastructure and power-supply facilities are often limited and the indoor environment changes as construction progresses, making localization systems that rely on dense wireless infrastructure or fingerprint databases difficult to deploy and maintain. Low-power wide-area (LPWA) communication can provide wide-area coverage with a small number of temporarily deployed access points (APs), but under sparse AP deployment, AP-to-terminal observations alone may provide insufficient floor-related information for reliable floor-level estimation.
To address this problem, this paper develops a device-centric cooperative
floor-level estimation method that exploits inter-terminal observations
as relative floor-difference information to complement AP-side information.
For practical application with limited measurement data, AP-to-terminal
observations are represented by an upper/lower-AP RSS contrast, while
inter-terminal observations are statistically modeled according to
floor-difference classes. The resulting AP-side and pairwise scores
are combined in a weighted floor-label objective and approximately
optimized using dual-order sequential iterated conditional modes
(ICM). Ray-tracing simulations and LoRa measurement experiments
demonstrate the effectiveness of the proposed method under
different AP deployments and terminal-selection limits.
\end{abstract}


\section{Introduction}
For construction-site applications in multi-story buildings, accurate
floor-level estimation is an important localization function for tracking
workers, construction equipment, and Internet of Things (IoT) devices~\cite{mostafa2025multifloor,li2020construction,chen2026constructionrobot}.
In buildings under construction, existing communication infrastructure
and power-supply facilities are often insufficient, while the indoor
environment continuously changes as construction progresses.
Accordingly, localization systems that rely on densely deployed wireless
infrastructure or fingerprint databases are difficult to deploy and maintain.
Such environments require localization techniques that can cover an entire
building using only a small number of temporarily deployed wireless devices.
This requirement is also consistent with the growing role of
localization and sensing in future wireless systems~\cite{kim2022isac_config,liu2022isac}.

Low-power wide-area (LPWA) technologies, especially LoRa, are attractive
for such environments because of their long communication range and low
power consumption~\cite{raza2017,augustin2016lora}. They can provide
wide-area coverage with a small number of temporarily deployed access
points (APs), which serve as anchors, while received signal strength (RSS)
can be obtained from ordinary packet reception without dedicated ranging
hardware or strict synchronization. 
Previous studies have investigated LoRa-based localization in both
indoor and outdoor environments~\cite{petajajarvi2015coverage,lam2019,li2018,chen2024}.
Floor-level estimation using unlicensed-band
LPWA signals was experimentally investigated in~\cite{nakano2022pimrc},
showing that observations from sparsely deployed APs contain useful
vertical information. However, when the number of APs is limited,
AP-to-terminal observations alone may not provide sufficient information
to reliably distinguish all floors.

One reason is the propagation behavior in multi-story buildings. RSS does
not necessarily provide a simple monotonic indication of floor separation
because indoor propagation is affected by floor penetration, exterior walls,
and surrounding structures~\cite{seidel1992pathloss,richalot2000,
kim2022building_entry_loss}. Ray-tracing studies have also shown that these
effects can considerably change floor-wise propagation characteristics~\cite{fuschini2015ray}. 
In~\cite{liang_jacecc_2025}, this propagation behavior was investigated through 
both ray-tracing simulations and measurement experiments and
explained using floor-transmitted (FT) and wall-transmitted reflected (WT)
components. The FT component decreases rapidly as the floor separation
increases because of cumulative penetration loss, whereas the WT component
can reach a receiver through reflected paths with less floor penetration.
Consequently, AP-side floor information becomes increasingly ambiguous as
the floor separation between an AP and a terminal increases, whereas
observations over smaller floor separations can remain more informative.

This observation motivates the use of device-centric cooperation based on 
the general concept of cooperative localization~\cite{patwari2005,
wymeersch2009,savarese2001locationing}. Our preliminary study
in~\cite{li_vtc2026} established a device-centric cooperative floor-level
estimation framework that combines AP-to-terminal and inter-terminal
observations. Although our preliminary study~\cite{li_vtc2026}
demonstrated the benefit of terminal cooperation under sparse
AP deployment, practical deployment requires reliable
floor-related observation models constructed from limited
site measurements. In real multi-story
environments, observations corresponding to different floors or floor
differences can overlap considerably, and parameter estimation errors under 
limited measurements may affect the overall floor-estimation performance.

In this paper, we propose a device-centric cooperative floor-level
estimation method for construction-site applications under sparse AP
deployment. Building on the cooperative floor-level estimation framework
developed in~\cite{li_vtc2026}, the proposed method extends the framework
toward practical measured environments by constructing the required
floor-related observation models from a limited number of site
measurements and introducing lower-complexity ICM-based inference.
The proposed method is evaluated through ray-tracing simulations
and LoRa measurements under different AP deployments and terminal-selection limits.
The main contributions of this paper are summarized as follows:
\begin{itemize}
\item A floor-related observation-model construction method using a limited
number of site measurements is developed for device-centric cooperative
floor-level estimation. The upper--lower AP RSS contrast provides AP-side
floor information, while inter-terminal observations provide statistical
information on relative floor differences.

\item The AP-side and pairwise scores are integrated into a
weighted floor-label objective on an OR-symmetrized
undirected graph and approximately optimized using
dual-order sequential iterated conditional modes (ICM).
Compared with BP, ICM reduces the per-link floor-state
processing from quadratic to linear dependence on the
number of candidate floors.

\item The proposed method is evaluated using ray-tracing simulations
and LoRa measurements under different AP deployments and
terminal-selection limits. The evaluation examines
floor-level estimation accuracy, floor-error distributions,
the validity of the AP-side observation model, and computation time.
\end{itemize}

\textit{Notation:}
Scalars are denoted by italic letters, vectors by bold lowercase
letters, and sets by calligraphic letters. The candidate floor-label set
is denoted by $\mathcal{K}=\{1,2,\ldots,K\}$, where $K$ is the number of
floors. The terminal floor label is denoted by $h_i\in\mathcal{K}$, and
the floor-label vector is
$\mathbf{h}=[h_1,h_2,\ldots,h_N]^{\mathsf T}\in\mathcal{K}^{N}$,
an $N\times 1$ integer-valued column vector. Its estimate
$\widehat{\mathbf h}$ has the same size.
The set of fixed APs is denoted by $\mathcal{A}$, and the terminal
collaboration graph by $\mathcal{G}=(\mathcal{V},\mathcal{E})$.
The notation $|\cdot|$ denotes set cardinality or absolute value
depending on the context. A hat indicates an estimate, and the
superscript $\star$ denotes a ground-truth label.

\section{Related Work}
\label{sec:related_work}

RSS-based indoor localization has been widely investigated because RSS
can be obtained from ordinary wireless communication without additional
ranging hardware~\cite{liu2007survey,zafari2019survey}.
Representative
fingerprinting approaches, such as RADAR and Horus, estimate location
by comparing measured RSS with a previously constructed radio
map~\cite{bahl2000radar,youssef2005horus}. 
Multi-floor indoor positioning has also been investigated using Wi-Fi
and multi-sensor observations~\cite{campos2014multifloor,li2018multisensor}.
Barometer-based methods infer floor levels from pressure
measurements using reference sensors or smartphone
barometers~\cite{xia2015barometer,ye2016barometer}.
In contrast, this study relies on wireless RSS observations.
Several studies have also investigated methods for reducing
the site-specific measurement
and calibration effort required for indoor localization~\cite{chintalapudi2010indoor,rai2012zee}.
Indoor localization and resource tracking at construction sites have also
been investigated using various wireless technologies~\cite{li2020construction}. 
However, studies specifically addressing floor-level estimation
under sparse, temporarily deployed infrastructure remain limited. For LPWA systems, \cite{nakano2022pimrc} experimentally
investigated floor-level estimation using sparsely deployed APs and
demonstrated the feasibility of exploiting floor-dependent received-signal
characteristics. The propagation characteristics relevant to multi-floor
environments were further investigated through both ray-tracing simulations 
and measurements in~\cite{liang_jacecc_2025},
providing insight into the floor-dependent behavior of received signals in
multi-story buildings.
Cooperative localization has been widely studied as a means of improving
position estimation by exploiting inter-node observations~\cite{patwari2005,wymeersch2009,savarese2001locationing}. In our
preliminary study~\cite{li_vtc2026}, this concept was introduced into
floor-level estimation by jointly using AP-to-terminal and inter-terminal
observations. Least-squares (LS) and Bayesian formulations were developed
using inter-terminal observations as relative floor-difference information,
with belief propagation (BP) used for approximate inference in the
Bayesian formulation.

Unlike the general indoor-localization and construction-site positioning
studies reviewed above, this paper specifically addresses device-centric
cooperative floor-level estimation under sparse AP deployment. Building
on our preliminary framework in~\cite{li_vtc2026}, the present work further
develops the AP-side and inter-terminal observation models using a limited
number of site measurements and introduces lower-complexity ICM-based
inference for evaluation in measured multi-story environments.

\section{System Model and Preliminary Formulation}
\label{sec:system_model}

\subsection{System Model}

\begin{figure}[t]
\centering
\hspace*{-5mm}
\includegraphics[width=0.9\linewidth]{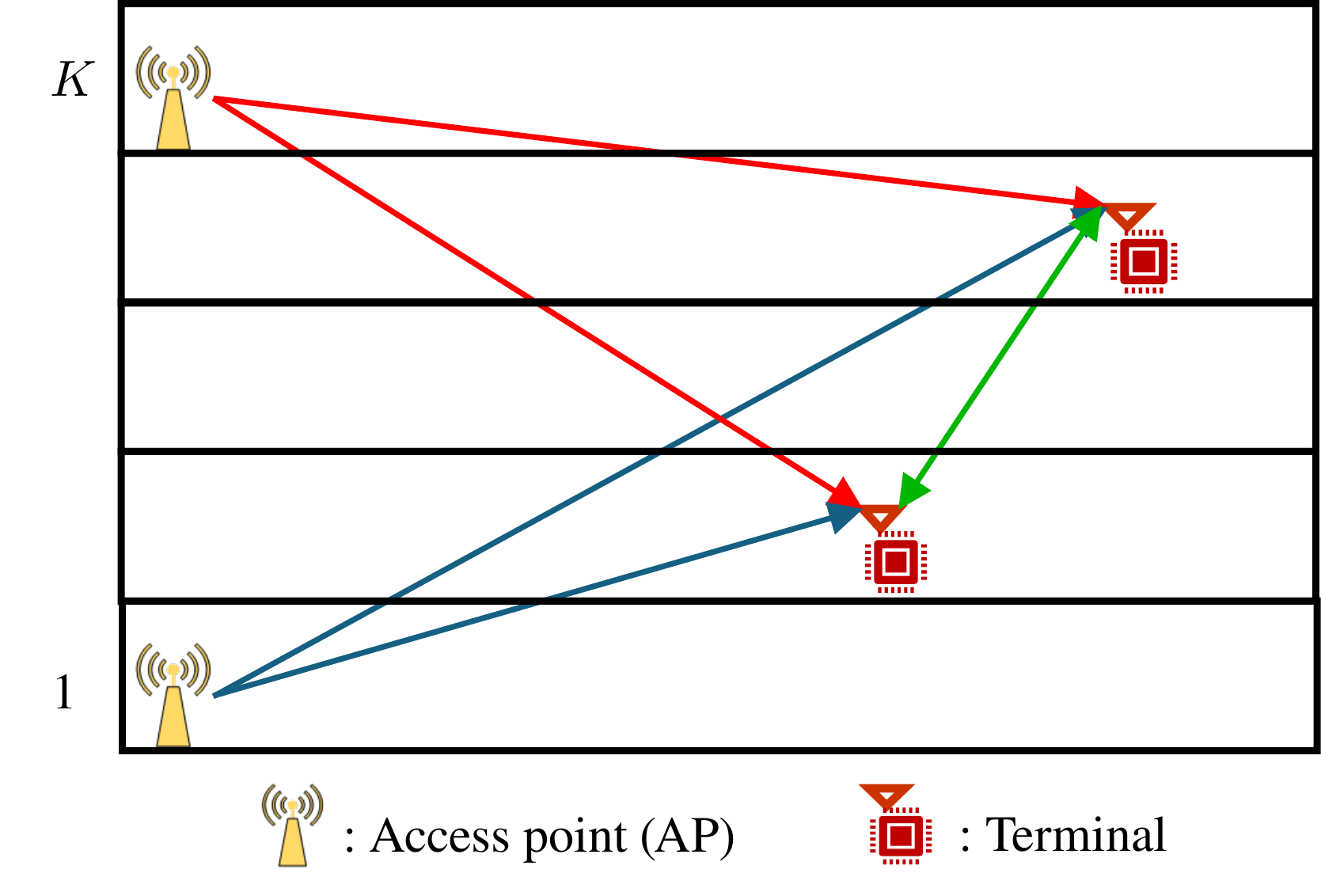}
\caption{System model for device-centric cooperative floor-level estimation under sparse AP deployment. Fixed APs provide AP-to-terminal RSS observations, while cooperating terminals provide inter-terminal RSS observations.}
\label{fig:system}
\end{figure}

Consider a multi-story building with $K$ floors, $M$ APs, and $N$
terminals, as illustrated in Fig.~\ref{fig:system}. The APs are
sparsely and temporarily deployed at fixed locations, and their floor
labels are known. APs need not be installed on every floor. The task is
to estimate the floor-label vector $\mathbf{h}$ of all terminals from
RSS observations.

The AP set is $\mathcal{A}=\{a_1,a_2,\ldots,a_M\}$.
Let $r_{i,a}\in\mathbb R$ denote the AP-to-terminal RSS
between terminal $i$ and AP $a$.
Terminal collaboration is represented by an undirected graph
$\mathcal{G}=(\mathcal{V},\mathcal{E})$,
where $\mathcal{V}=\{1,2,\ldots,N\}$.
For a terminal-selection limit $L\in\{0,1,\ldots,N-1\}$,
each terminal $i$ independently selects up to $L$ other
terminals in descending order of inter-terminal RSS.
Let $\mathcal{S}_i^{\mathrm{sel}}\subseteq\mathcal V\setminus\{i\}$
denote the set of terminals selected by terminal $i$.
Define the binary indicator $A_{ij}\in\{0,1\}$ by $A_{ij}=1$ if
$j\in\mathcal{S}_i^{\mathrm{sel}}$ and $A_{ij}=0$ otherwise.

The undirected edge set is constructed as
\begin{equation}
\mathcal{E} =
\left\{\{i,j\}\mid A_{ij}=1 \lor A_{ji}=1\right\}.
\end{equation}
The set of terminals connected to terminal $i$ is
$\mathcal{N}_i=\{j\mid\{i,j\}\in\mathcal{E}\}$.
Thus, $L$ limits the number of terminals actively selected
by each terminal, whereas the degree of a terminal in the
undirected graph may exceed $L$.
The setting $L=0$ disables cooperation, whereas $L=N-1$
allows each terminal to select all other terminals with
available inter-terminal RSS observations before OR
symmetrization.

\subsection{LS and Bayesian Formulations in the Preliminary Study}

Our preliminary study~\cite{li_vtc2026} formulated floor-level
estimation under sparse AP deployment using AP-to-terminal and
inter-terminal observations. In the following expressions,
$\mathcal E_{\rm pre}$ denotes the terminal-link set of the respective
preliminary method; it need not coincide with the OR-symmetrized
edge set $\mathcal E$ used by the proposed method.
The AP-side information for terminal $i$ is represented either by
an AP-only floor estimate $\widetilde h_i^{\rm AP}\in\mathcal K$
or by a prior mass function $p_i^{\rm AP}(k)$ over
$k\in\mathcal K$, while the inter-terminal RSS provides
relative information about the floor difference $|h_i-h_j|$.

In the LS realization, the floor labels are estimated by minimizing
the inconsistency between the estimated floor differences and the
floor-difference observations obtained from inter-terminal RSS:
\begin{equation}
\begin{aligned}
J_{\rm LS}(\mathbf h)
={}&\sum_{i=1}^{N}\lambda_i
\left(h_i-\widetilde h_i^{\rm AP}\right)^2\\
&+\sum_{(i,j)\in\mathcal E_{\rm pre}}w_{ij}
\left(\widehat d_{ij}-|h_i-h_j|\right)^2,
\end{aligned}
\label{eq:base_ls}
\end{equation}
where $\widehat d_{ij}\in\mathbb R$ is a scalar floor-difference
estimate derived from the inter-terminal RSS. The real-valued
scalar weights $\lambda_i$ and $w_{ij}$ control the AP-side
and pairwise terms, respectively. 

The Bayesian realization represents the same cooperative information
probabilistically. Its MAP formulation can be written as
\begin{equation}
\begin{aligned}
\widehat{\mathbf h}=
\argmax_{\mathbf h\in\mathcal K^N}\Bigl[
&\sum_{i=1}^{N}\log p_i^{\rm AP}(h_i)\\
&+\sum_{(i,j)\in\mathcal E_{\rm pre}}
\log p(r_{ij}\mid |h_i-h_j|)
\Bigr].
\end{aligned}
\label{eq:base_bayesian}
\end{equation}
Belief propagation (BP) is used for approximate inference in this
formulation. These LS and Bayesian realizations are referred to as
Coop-LS and Coop-BP, respectively.

\section{Proposed Cooperative Floor-Level Estimation}
\label{sec:proposed}
\begin{figure}[t]
\centering
\includegraphics[width=0.95\linewidth]{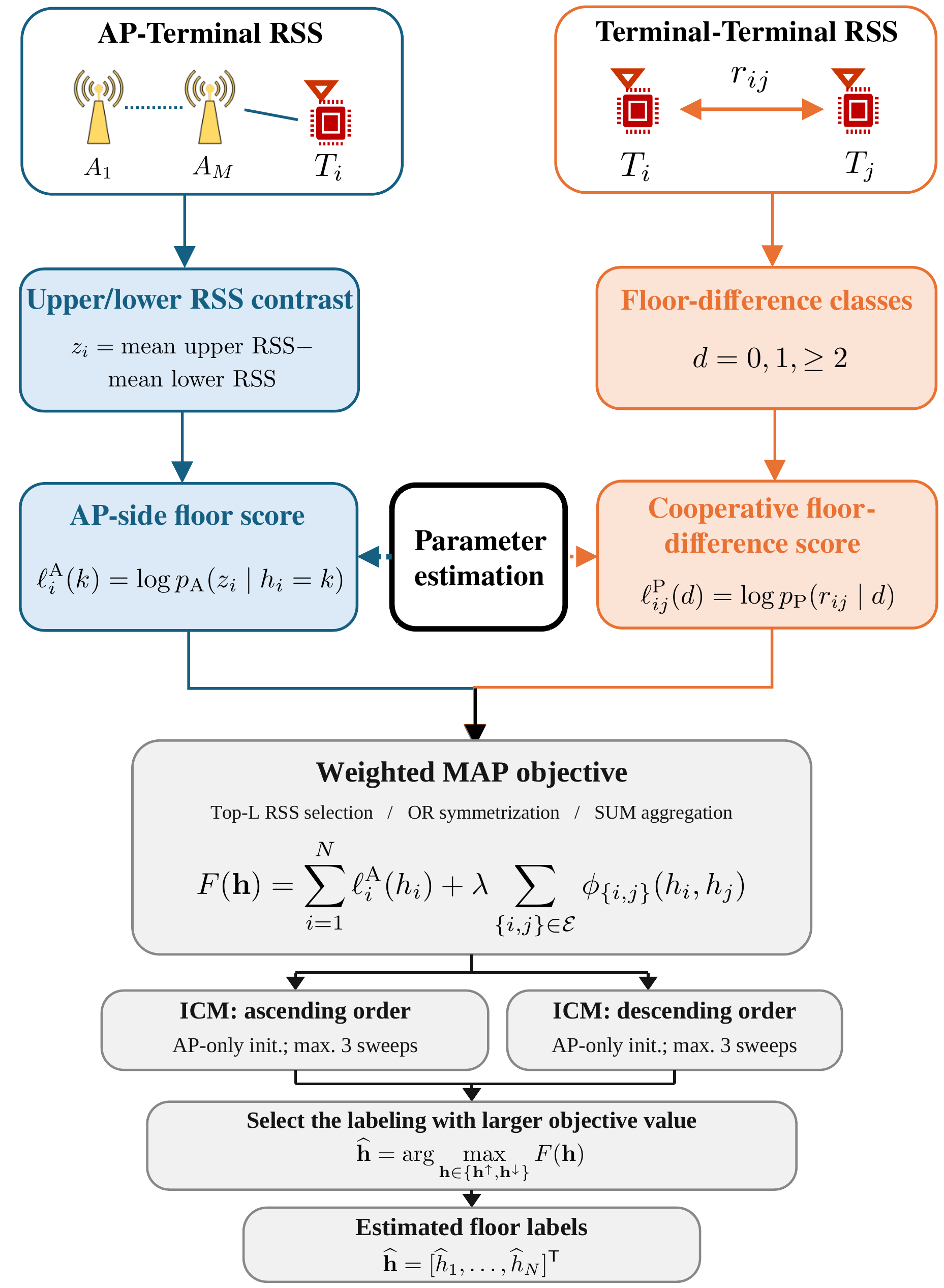}
\caption{Overview of the proposed floor-level estimation method.
AP-side and inter-terminal RSS are converted into statistical scores,
which are combined through an undirected pairwise objective. Two
sequential ICM runs with opposite update orders generate candidate
labelings, and the candidate with the higher objective value is selected.}
\label{fig:observation}
\end{figure}

Building on Section~\ref{sec:system_model}, the proposed method
constructs AP-side and pairwise observation models from limited site
measurements. Their scores are combined in a weighted floor-label
objective on an undirected graph, and the labels are estimated using
dual-order sequential ICM, as summarized
in Fig.~\ref{fig:observation}.

\subsection{Proposed Algorithm}


\noindent
\textbf{1) AP-side floor score:}

We consider a sparse AP deployment in which APs are placed on the top
and bottom floors of the building, while no APs are required on the
intermediate floors. The relative strength of the two AP groups provides
floor-related information for terminals located throughout the building.
Let $\mathcal{A}_{\rm U},\mathcal{A}_{\rm L}\subseteq\mathcal A$
denote the nonempty AP sets on the top and bottom floors,
respectively. For terminal $i$, the mean RSS values
$\bar r_{i,\rm U},\bar r_{i,\rm L}\in\mathbb R$ are
\begin{align}
\bar r_{i,\rm U}
&=\frac{1}{|\mathcal{A}_{\rm U}|}
\sum_{a\in\mathcal{A}_{\rm U}}r_{i,a}, &
\bar r_{i,\rm L}
&=\frac{1}{|\mathcal{A}_{\rm L}|}
\sum_{a\in\mathcal{A}_{\rm L}}r_{i,a}.
\end{align}

We use the difference between these two averages as a compact
AP-side observation $z_i\in\mathbb R$:
\begin{equation}
z_i=\bar r_{i,\rm U}-\bar r_{i,\rm L}.
\label{eq:ap_contrast}
\end{equation}
This contrast reduces the influence of RSS variations common to the two
AP groups and emphasizes the change in their relative signal strengths
along the vertical direction.

To represent the AP-side floor information probabilistically, the score
for candidate floor $k$ is defined as
\begin{equation}
\ell_i^{\rm A}(k)
=
\log p_{\rm A}(z_i\mid h_i=k),
\label{eq:ap_score}
\end{equation}
where $p_{\rm A}(z_i\mid h_i=k)$ denotes a conditional Gaussian
probability density and $\ell_i^{\rm A}(k)\in\mathbb R$ is the
corresponding log-density score. This probabilistic representation is consistent with the Bayesian
formulation in our preliminary study~\cite{li_vtc2026}, while the
AP-side observation model itself is constructed here from a limited
number of site measurements.

When the number of measurement samples is limited, the directly
estimated floor-wise means may not preserve the expected monotonic
relationship with the floor index. We therefore estimate the floor-wise
means from the available measurements and apply isotonic regression to
restore their monotonic order. 
The variance is lower-bounded by $\sigma_{\mathrm{A,min}}^2$
to prevent an excessively sharp likelihood.
Once determined, these model parameters are fixed throughout
the subsequent evaluation. The ability of the resulting AP-side model
to reproduce the broader floor-wise RSS-contrast distributions is
examined experimentally in Section~\ref{sec:calibration_validation}.


\noindent\textbf{2) Cooperative floor-difference score:}

The AP-side score may be ambiguous, particularly for intermediate
floors. Inter-terminal RSS supplies complementary information about
relative floor separation. For an ordered observation from terminal
$i$ to terminal $j$, let $r_{ij}\in\mathbb R$ denote the
measured RSS; its conditional distribution is
approximated as a function of floor
difference:
\begin{equation}
p(r_{ij}\mid h_i,h_j)
\simeq
p(r_{ij}\mid |h_i-h_j|).
\label{eq:pairwise_approximation}
\end{equation}

In measured multi-story environments, RSS distributions corresponding
to large floor separations can overlap substantially. Distinguishing
every possible floor separation independently may therefore result in
poorly characterized observation classes. We consequently group the
floor difference as
\begin{equation}
d_{ij}=\min(|h_i-h_j|,2),
\qquad d_{ij}\in\{0,1,2\},
\label{eq:floor_difference}
\end{equation}
where $d_{ij}=2$ represents a floor difference of two or more floors.
The pairwise score for class $d$ is defined as
\begin{equation}
\ell_{ij}^{\rm P}(d)
=
\log p_{\rm P}(r_{ij}\mid d),
\label{eq:pairwise_score}
\end{equation}
where $d\in\{0,1,2\}$ and $p_{\rm P}(r_{ij}\mid d)$ denotes
a conditional Gaussian probability density. The pairwise
score $\ell_{ij}^{\rm P}(d)$ is real-valued.
The variance of this distribution is lower-bounded by $\sigma_{\mathrm{P,min}}^2$ 
to prevent an excessively sharp likelihood. The class parameters are 
estimated from a limited number of terminal-to-terminal measurements and 
are then fixed throughout the subsequent evaluation.

\noindent\textbf{3) MAP formulation and ICM-based inference:}

The AP-side score represents the floor evidence for each terminal,
while the pairwise score measures consistency with the relative
floor differences inferred from connected terminals.
For each undirected edge $\{i,j\}\in\mathcal{E}$ and
candidate labels $a,b\in\mathcal K$, we define the real-valued
aggregated pairwise score as
\begin{equation}
\begin{aligned}
\phi_{\{i,j\}}(a,b)
={}& A_{ij}\ell_{ij}^{\rm P}
\left(\min(|a-b|,2)\right)\\
&+A_{ji}\ell_{ji}^{\rm P}
\left(\min(|a-b|,2)\right).
\end{aligned}
\label{eq:pair_potential}
\end{equation}

The accumulated scores are
\begin{align}
J_{\rm A}(\mathbf h)
&=\sum_{i=1}^{N}\ell_i^{\rm A}(h_i),\\
J_{\rm P}(\mathbf h)
&=\sum_{\{i,j\}\in\mathcal E}
\phi_{\{i,j\}}(h_i,h_j).
\end{align}

The real-valued weighted objective $F:\mathcal K^N\to\mathbb R$ is
\begin{equation}
F(\mathbf h)=
J_{\rm A}(\mathbf h)+\lambda J_{\rm P}(\mathbf h).
\label{eq:joint_metric}
\end{equation}

The nonnegative real-valued weight $\lambda$ balances AP-side and
inter-terminal evidence. Under a uniform prior
over floor labelings, the weighted objective
can be interpreted as a surrogate log-posterior,
without requiring the pairwise RSS observations
to be statistically independent.
For the proposed method, the empirically selected weight is fixed
at $\lambda=0.05$ in both ray-tracing scenarios and the LoRa experiment.


An exhaustive search of~\eqref{eq:joint_metric} is impractical because
the number of possible floor-label combinations grows as $K^N$.
We therefore use ICM~\cite{besag1986dirty} for approximate inference.
Each terminal is first assigned to the
floor that maximizes its AP-side score.
With the current labels of the connected terminals fixed,
the real-valued local score $S_i(k)$ of terminal $i$ for
candidate floor $k\in\mathcal K$ is
\begin{equation}
S_i(k)=
\ell_i^{\rm A}(k)
+\lambda\sum_{j\in\mathcal N_i}
\phi_{\{i,j\}}(k,h_j).
\label{eq:local_score}
\end{equation}

Two sequential ICM procedures are independently performed
from the same AP-only initialization.
One procedure updates terminals in ascending terminal-index order,
whereas the other uses descending terminal-index order.

Within each procedure, updated labels are immediately
used in subsequent coordinate updates.
If the current label is among the maximizers,
it is retained to avoid unnecessary changes.

Each procedure terminates when a complete sweep produces
no label changes or after a maximum of three sweeps.
Let $\mathbf h^{\uparrow},\mathbf h^{\downarrow}\in\mathcal K^N$
denote the resulting $N\times 1$ floor-label vectors.

The final estimate is selected according to
\begin{equation}
\widehat{\mathbf h}=
\argmax_{\mathbf h\in
\{\mathbf h^{\uparrow},\mathbf h^{\downarrow}\}}
F(\mathbf h).
\label{eq:dual_order_selection}
\end{equation}

Each sequential coordinate update does not decrease $F$,
and the higher-scoring final labeling is selected.

\subsection{Computational Complexity}

AP-side score construction requires processing $M$ AP observations for
each of the $N$ terminals. The selected graph and likelihood scores
are computed once and shared by both ICM runs. Let $E=|\mathcal E|$,
$R=2$ be the number of update orders, and $T_{\max}=3$ the maximum
number of sweeps per order. With the graph and likelihood scores
precomputed, the dual-order ICM inference cost is
\begin{equation}
\mathcal{O}\!\left(RT_{\max}K(N+E)\right).
\label{eq:proposed_complexity}
\end{equation}
Online prediction also includes AP score calculation,
Top-$L$ terminal selection, graph construction, and pairwise-score
precomputation.

The BP realization in our preliminary study~\cite{li_vtc2026} has
complexity
\begin{equation}
\mathcal{O}\!\left(T_{\rm BP}E_{\rm pre}K^2\right).
\label{eq:bp_complexity}
\end{equation}
Here, $T_{\rm BP}$ denotes the number of BP iterations,
and $E_{\rm pre}=|\mathcal E_{\rm pre}|$
is the number of links in the preliminary graph. Each BP message
update evaluates pairs of $K$-state labels, whereas ICM evaluates
$K$ candidate labels with the labels of connected terminals fixed.
Thus, the per-link state dependence is $\mathcal{O}(K^2)$ for BP and
$\mathcal{O}(K)$ for ICM. Fig.~\ref{fig:runtime} compares the measured
online prediction times of the complete methods.

\section{Simulation Evaluation}
\label{sec:simulation_evaluation}

\subsection{Simulation Setup}

\begin{figure}[t]
    \centering
    \subfloat[Simplified building model]{
        \includegraphics[width=0.9\linewidth]{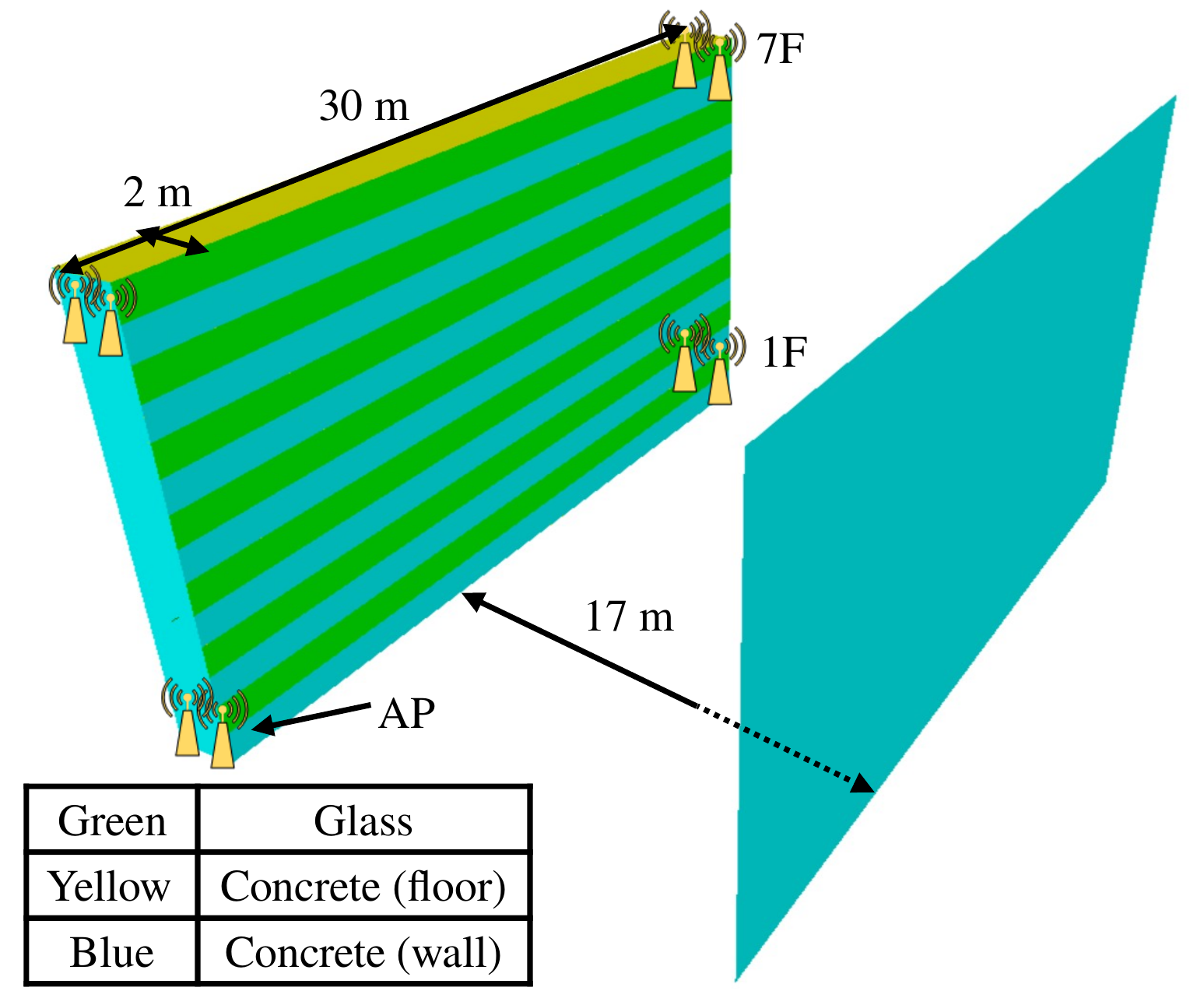}
        \label{fig:simulation_normal}
    }
    \vspace{1mm}
    \subfloat[Realistic building model]{
        \includegraphics[width=0.9\linewidth]{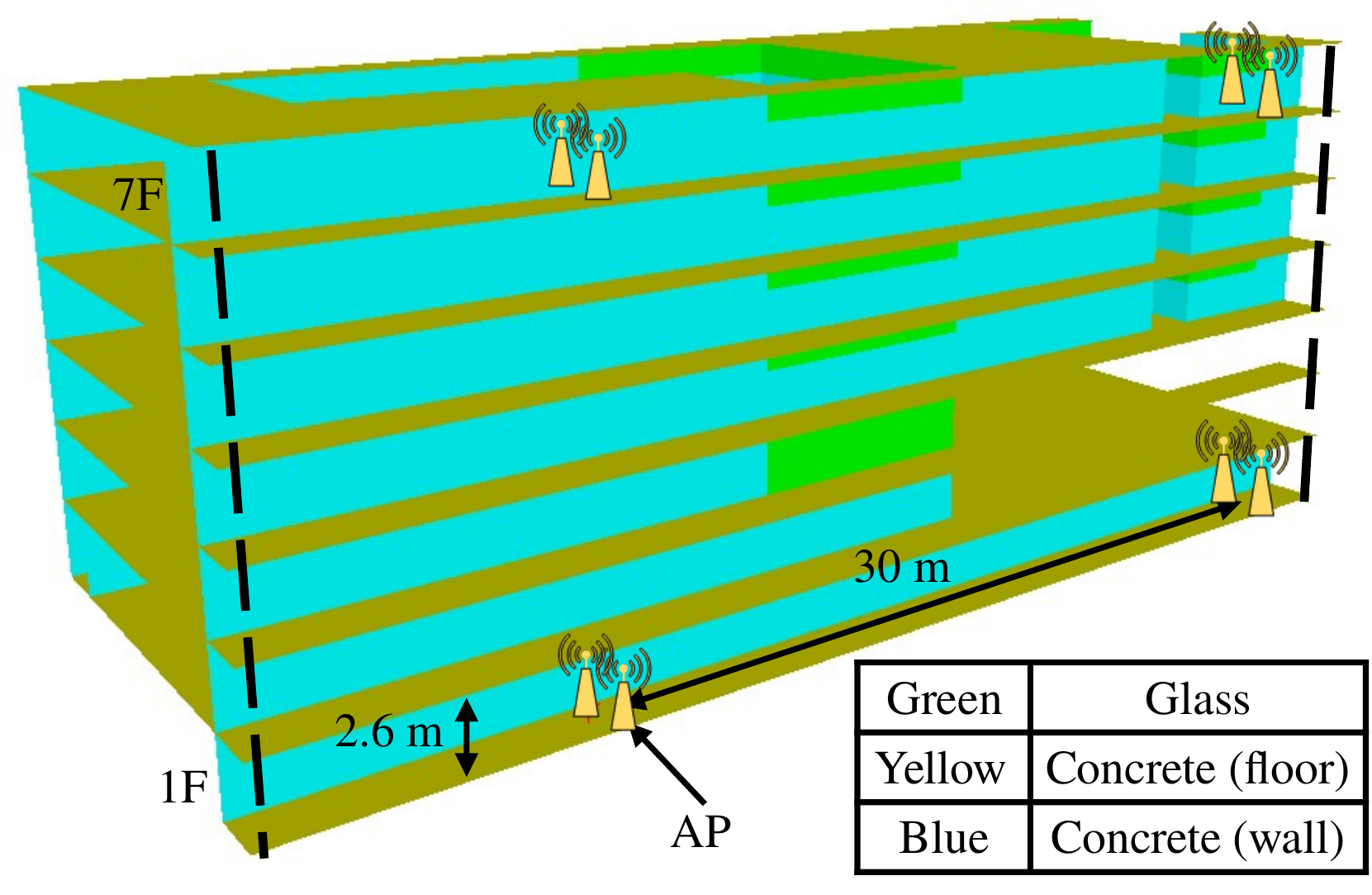}
        \label{fig:simulation_west2}
    }
    \caption{Simulation models used for ray-tracing evaluation:
    (a) simplified building model with a seven-floor corridor and an
    exterior reflection wall, and (b) realistic building model reproducing
    the corridor and atrium geometry of the experimental environment.}
    \label{fig:sim_models}
\end{figure}

Ray-tracing simulation is used to evaluate the methods under controlled
multi-floor propagation conditions. Fig.~\ref{fig:sim_models}(a) shows
a simplified building model consisting of a seven-floor corridor with
an exterior reflection wall. Fig.~\ref{fig:sim_models}(b) shows a
realistic building model that more closely reproduces the corridor and
atrium geometry of the experimental environment. The two models enable
the effects of scene-dependent reflection, wall transmission, and floor
transmission to be examined without short-term measurement fluctuations.

In the simplified building model, six terminal positions are randomly
selected along the corridor centerline on each floor in each of ten
simulation trials. Floor-estimation accuracy is averaged over these
trials. In the realistic building model, six fixed terminal positions
per floor (42 positions in total) are used to match the positions in
the LoRa measurement experiment.
Under AP4+4, four APs are deployed on each of the first and seventh
floors. Under AP1+1, one AP is deployed on each of these floors to
evaluate a sparser AP deployment. The carrier frequency is 920.7\,MHz
for both AP-to-terminal and terminal-to-terminal links.
In the simulations, the AP-to-terminal and inter-terminal RSS values
supplied to the observation models are expressed in dBm, and the
resulting upper--lower AP RSS contrast is expressed in dB.
The terminal-selection limit $L$, defined as the maximum number of
other terminals actively selected by each terminal, is varied over
$L\in\{0,1,3,5,10,20,30,N-1\}$ ($N-1=41$).
For both AP deployments, the proposed method independently performs 
sequential ICM in ascending and descending terminal-index orders, 
starting from the same AP-only initialization. Each run terminates 
after at most three complete sweeps, with early stopping if a sweep 
produces no label changes. The final estimate is selected by comparing 
the weighted objective values of the two resulting candidates.
The variance lower bounds are set to $\sigma_{\mathrm{A,min}}^2=1\,\mathrm{dB}^2$
for the AP-side RSS contrast and $\sigma_{\mathrm{P,min}}^2=9\,\mathrm{dB}^2$
for inter-terminal RSS.

The proposed method is compared with Coop-LS and Coop-BP, which denote
the LS and Bayesian/BP realizations of our preliminary cooperative
framework in~\cite{li_vtc2026}, respectively. 
AP-only, based on the floor-level estimation method in~\cite{nakano2022pimrc}, 
is included as a noncooperative reference. All methods use the same AP deployments and terminal-selection
limit $L$. However, the proposed method uses an OR-symmetrized graph,
whereas Coop-LS and Coop-BP retain their respective
preliminary link constructions.

The floor-wise accuracy is defined as the fraction of correctly
estimated terminals among those located on each ground-truth floor,
while the overall accuracy is the fraction of correctly estimated
terminals over all evaluated positions. To characterize not only exact
floor accuracy but also the distribution of estimation errors, we also
use the predicted-floor probability mass function (PMF), defined as the
empirical distribution of estimated floor labels for each ground-truth
floor.

\subsection{Floor-Wise Accuracy in the Two Building Models}

\begin{figure}[t]
    \centering
    \subfloat[Simplified building model]{
        \includegraphics[width=0.9\linewidth]{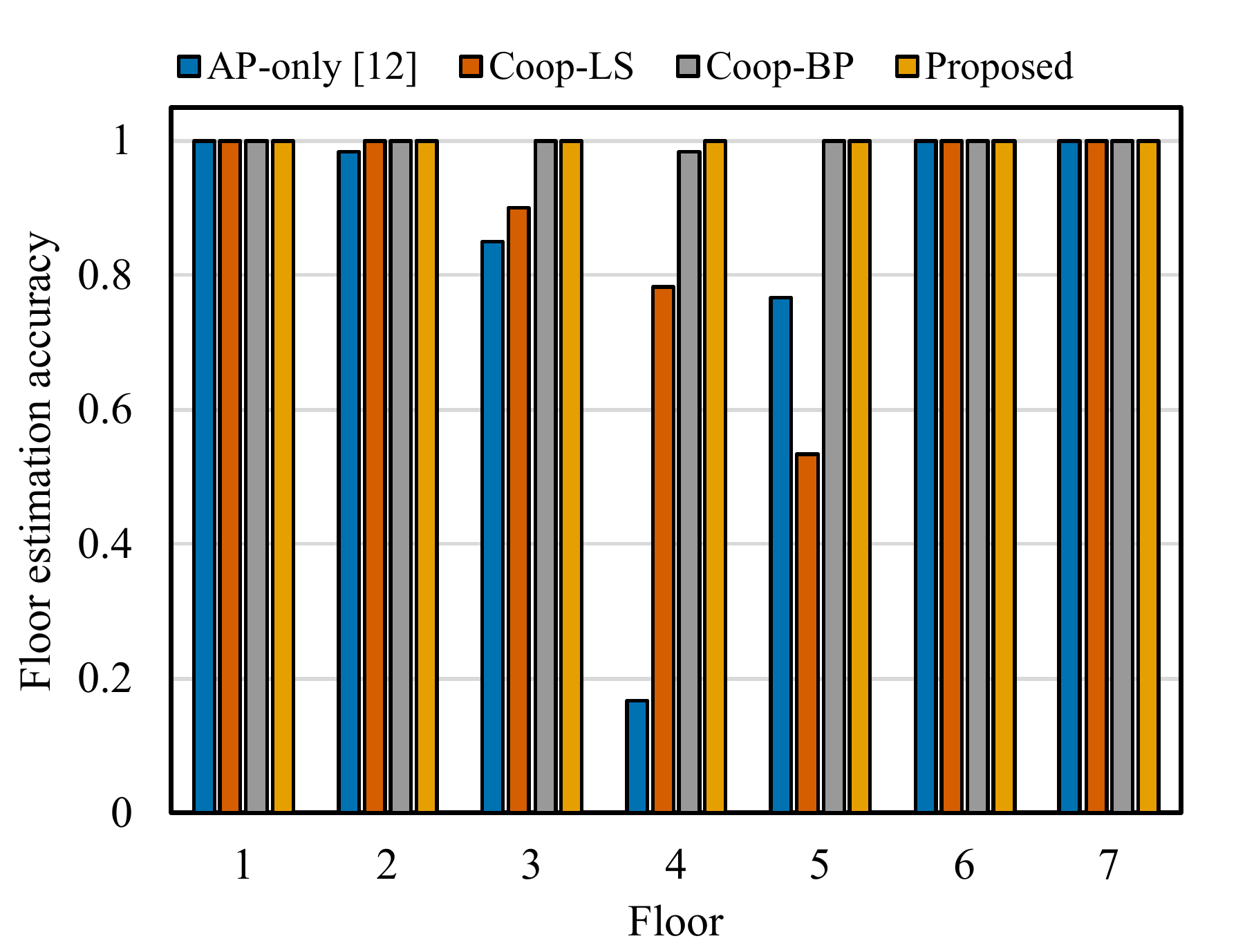}
        \label{fig:sim_floorwise_normal}
    }
    \vspace{-3mm}
    \subfloat[Realistic building model]{
        \includegraphics[width=0.9\linewidth]{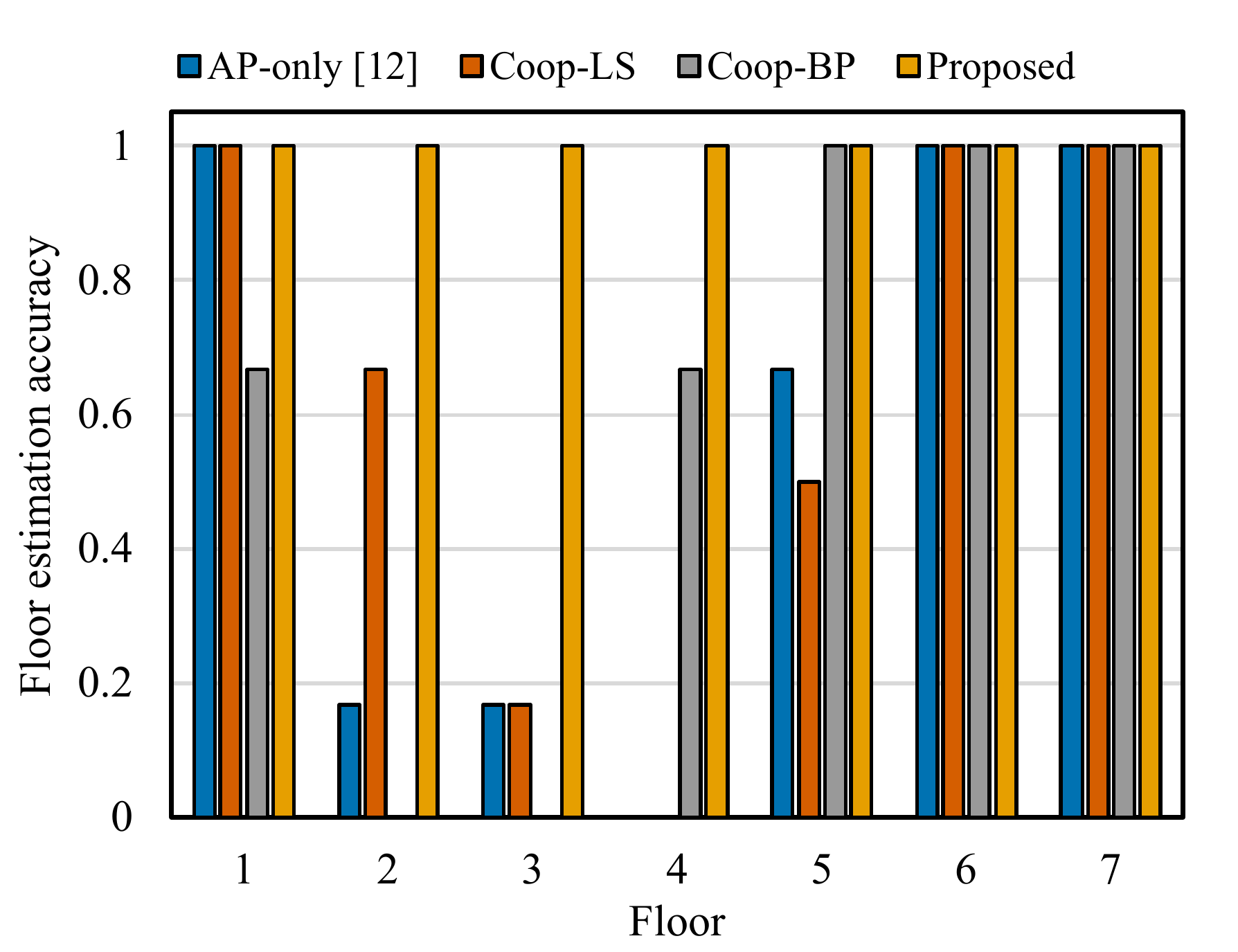}
        \label{fig:sim_floorwise_west2}
    }
    \caption{Floor-wise estimation accuracy under AP4+4
    with $L=N-1$ in the ray-tracing simulations:
    (a) simplified building model, averaged over ten trials
    with six randomly selected terminal positions per floor,
    and (b) realistic building model with six fixed terminal
    positions per floor corresponding to the LoRa experiment.}
    \label{fig:sim_floorwise}
\end{figure}

Fig.~\ref{fig:sim_floorwise} compares floor-wise accuracy under AP4+4.
In the simplified building model, the RSS-to-floor-difference
relationship is relatively regular, and the cooperative methods in our
preliminary study can effectively exploit terminal-to-terminal RSS.
The proposed method preserves this cooperative gain using the revised
statistical observation models. In the realistic building model, the
more complex propagation geometry results in stronger overlap among
floor-dependent RSS patterns. Under this condition, performance depends
not only on whether terminal cooperation is used, but also on how the
AP-side and pairwise observations are represented as floor-related
scores. The stable floor-wise performance of the proposed method
indicates the benefit of representing ambiguous RSS observations using
statistical scores rather than directly converting RSS into a
floor-related quantity.

\subsection{Simplified Building Model:
PMFs and Terminal Selection}

\begin{figure}[t]
    \centering
    \subfloat[AP4+4]{
        \includegraphics[width=0.85\linewidth]{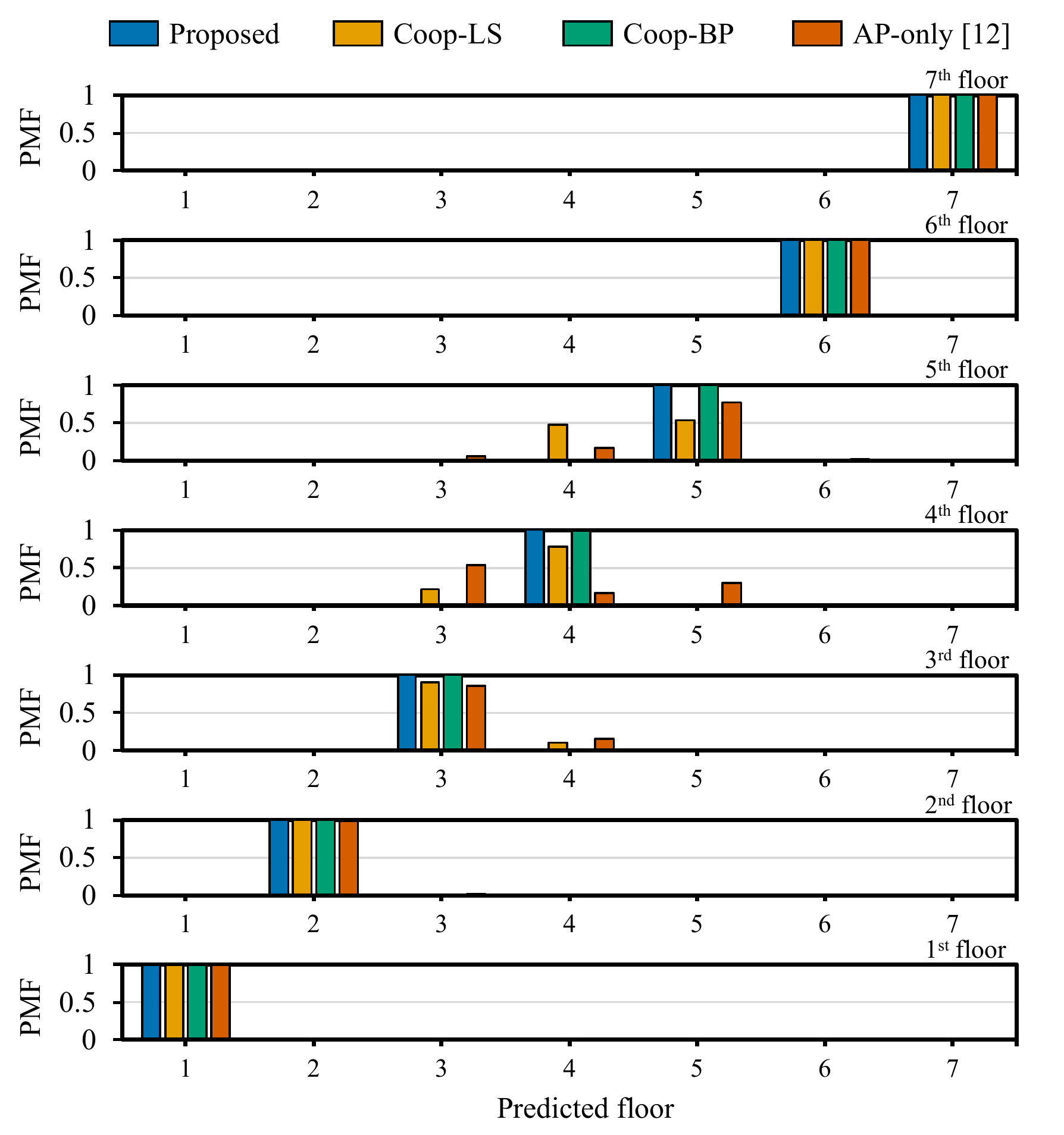}
        \label{fig:normal_pmf_ap44}
    }
    \vspace{-3mm}
    \subfloat[AP1+1]{
        \includegraphics[width=0.85\linewidth]{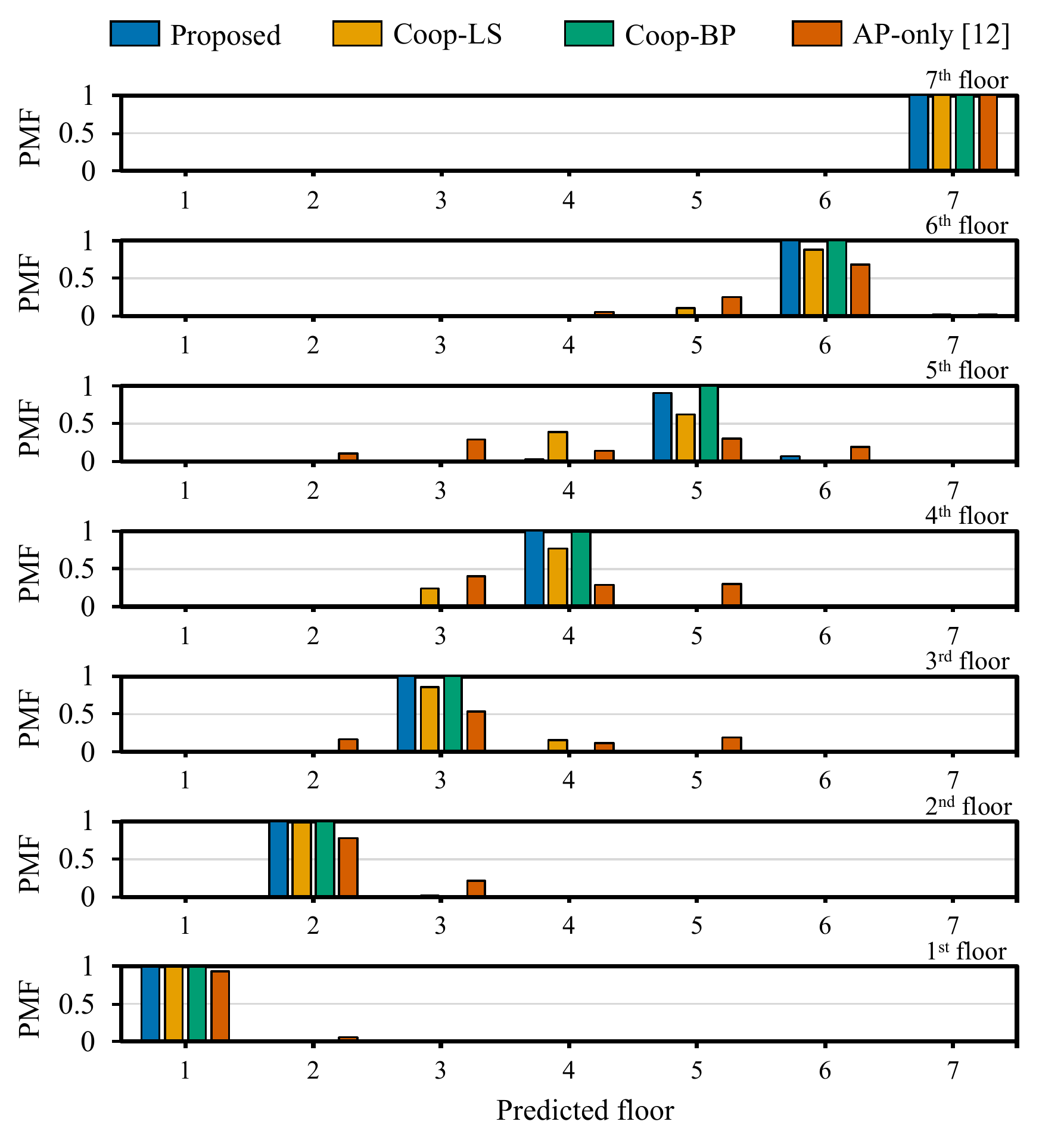}
        \label{fig:normal_pmf_ap11}
    }
    \caption{Predicted-floor PMFs in the simplified building
    model with $L=N-1$, aggregated over ten trials with
    six randomly selected terminal positions per floor:
    (a) AP4+4 and (b) AP1+1.
    Each row corresponds to a true floor, and the bars show
    the estimated-floor distribution for each method.}
    \label{fig:normal_pmf}
\end{figure}

Fig.~\ref{fig:normal_pmf} shows the predicted-floor PMFs 
aggregated over the ten simulation trials under AP4+4 and AP1+1. 
A narrow distribution centered on the ground-truth floor
indicates both high estimation accuracy and small floor-estimation
errors. Under AP4+4, the APs on the top and bottom floors provide
strong floor-related information, and terminal cooperation mainly
refines estimates on the more ambiguous intermediate floors. Under
AP1+1, AP-side floor discrimination becomes less reliable, increasing
the relative contribution of terminal-to-terminal information. The
PMFs also show whether estimation errors remain close to the
ground-truth floor or shift toward more distant or boundary floors.

\begin{figure}[t]
    \centering
    \subfloat[AP4+4]{
        \includegraphics[width=0.9\linewidth]{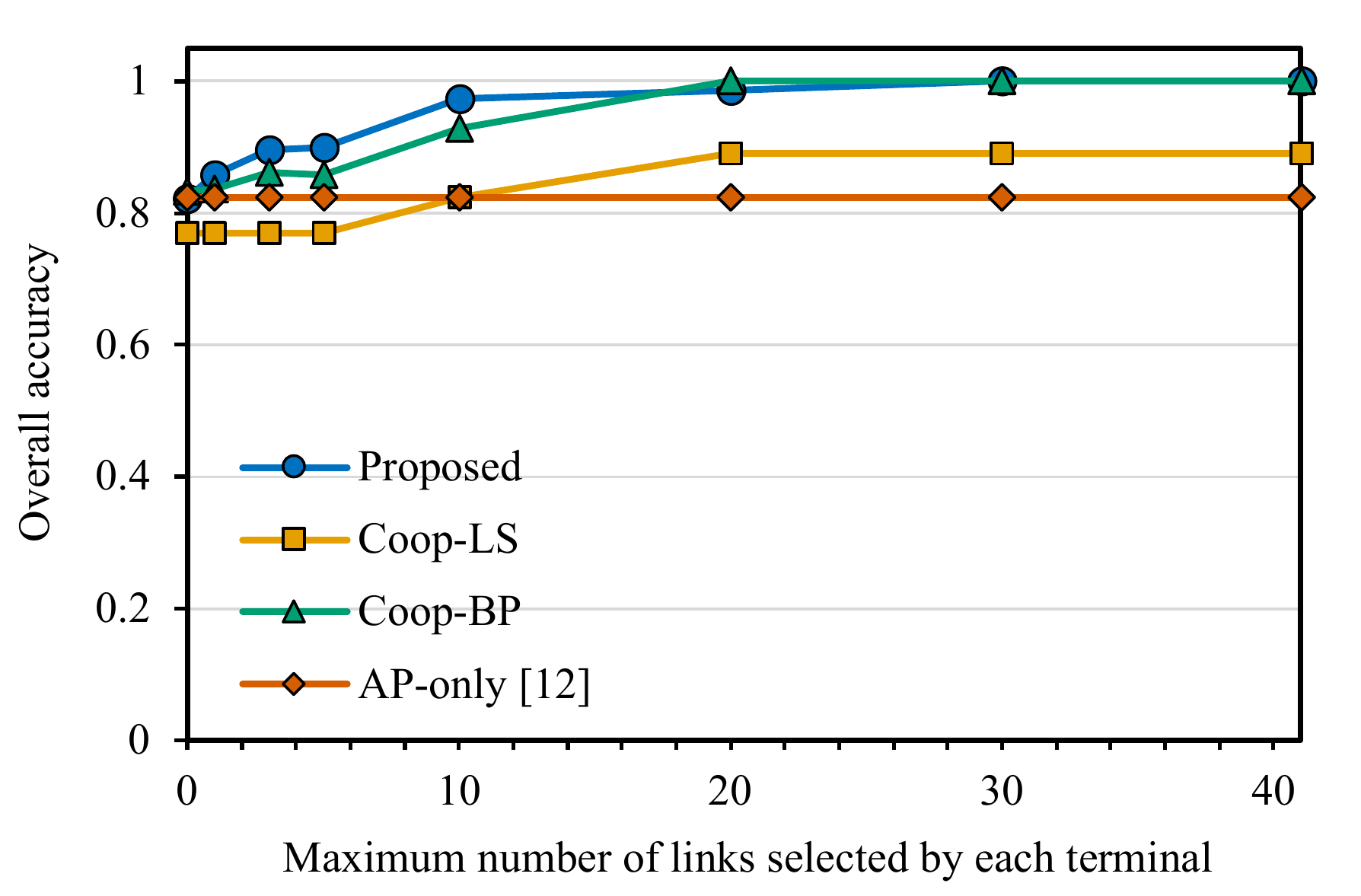}
        \label{fig:normal_links_ap44}
    }
    \vspace{-3mm}
    \subfloat[AP1+1]{
        \includegraphics[width=0.9\linewidth]{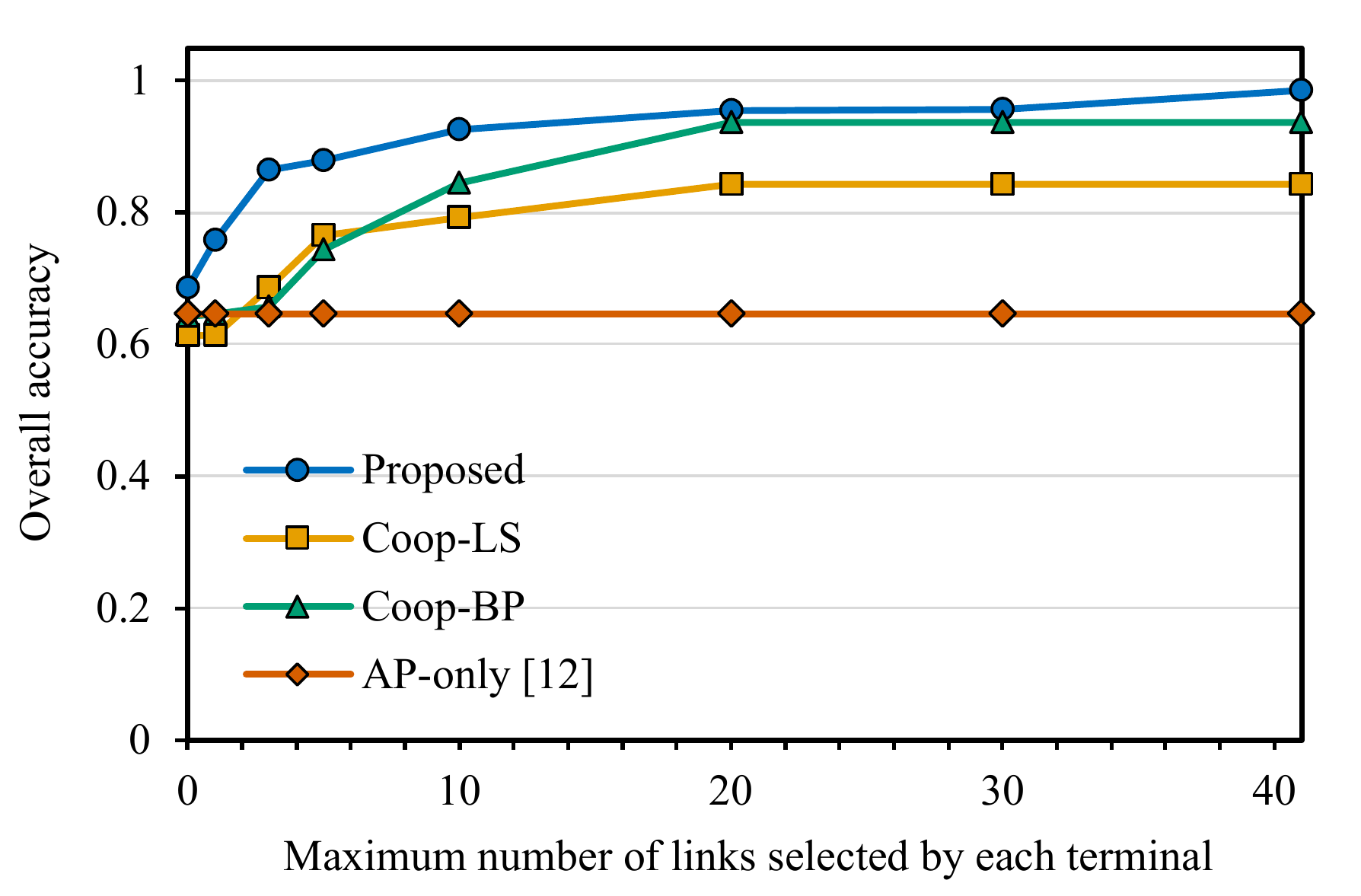}
        \label{fig:normal_links_ap11}
    }
    \caption{Overall floor-estimation accuracy versus
    terminal-selection limit $L$ in the simplified building
    model, averaged over ten trials with six randomly
    selected terminal positions per floor:
    (a) AP4+4 and (b) AP1+1.
    AP4+4 uses four APs on each of floors 1 and 7;
    AP1+1 uses one AP on each of these floors.}
    \label{fig:normal_links}
\end{figure}

Fig.~\ref{fig:normal_links} shows overall accuracy across
terminal-selection limits. At $L=0$, each method uses
only its AP-side information.
The proposed method retains high accuracy over a broad
range of selection limits, indicating that cooperative gains
can be maintained even when each terminal selects only
a subset of the other terminals with available RSS observations.
Under AP1+1, the more limited AP-side floor information makes
performance more sensitive to the terminal-selection limit.

\subsection{Realistic Building Model:
PMFs and Terminal Selection}

\begin{figure}[t]
    \centering
    \subfloat[AP4+4]{
        \includegraphics[width=0.85\linewidth]{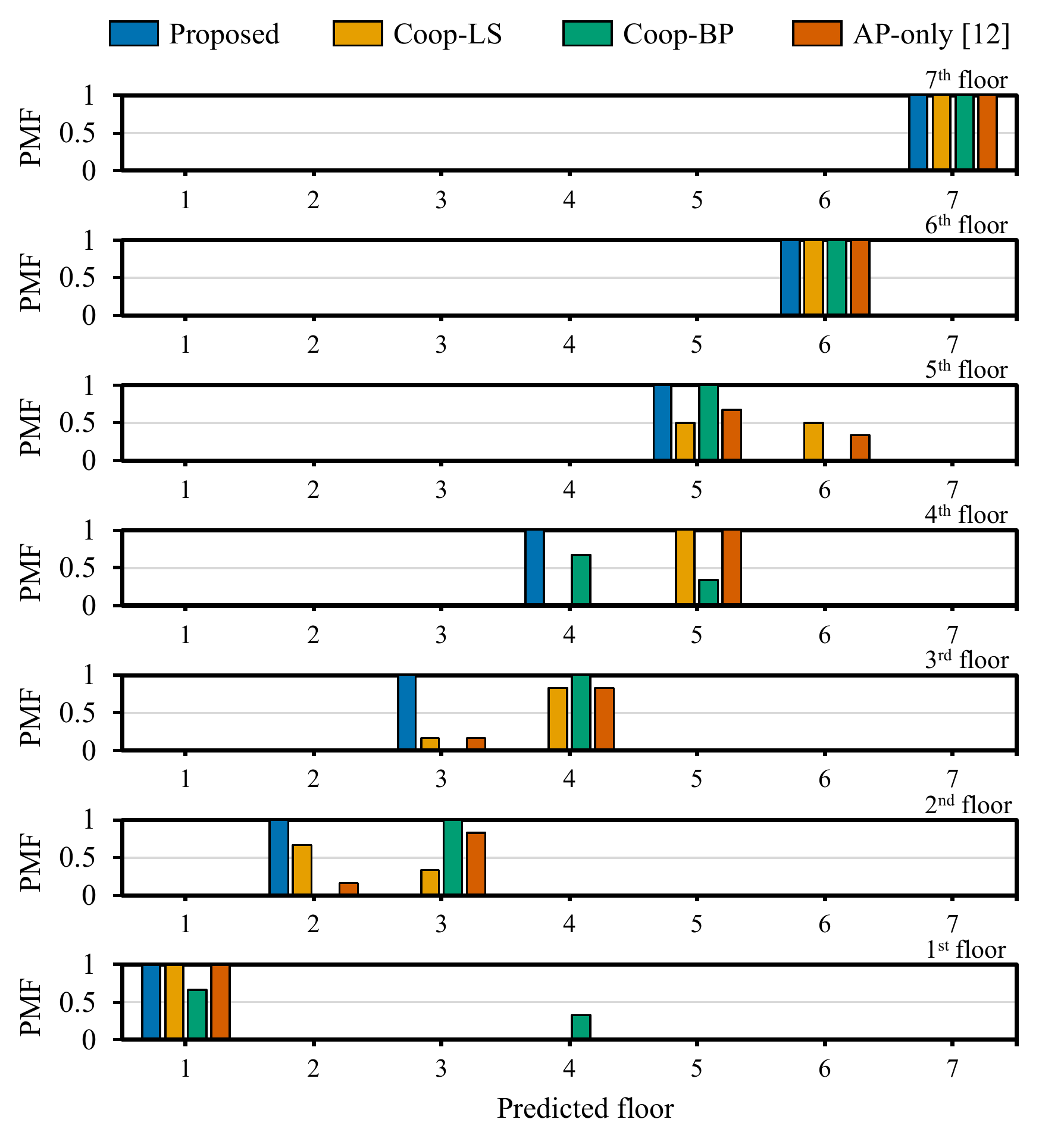}
        \label{fig:west2_pmf_ap44}
    }
    \vspace{-3mm}
    \subfloat[AP1+1]{
        \includegraphics[width=0.85\linewidth]{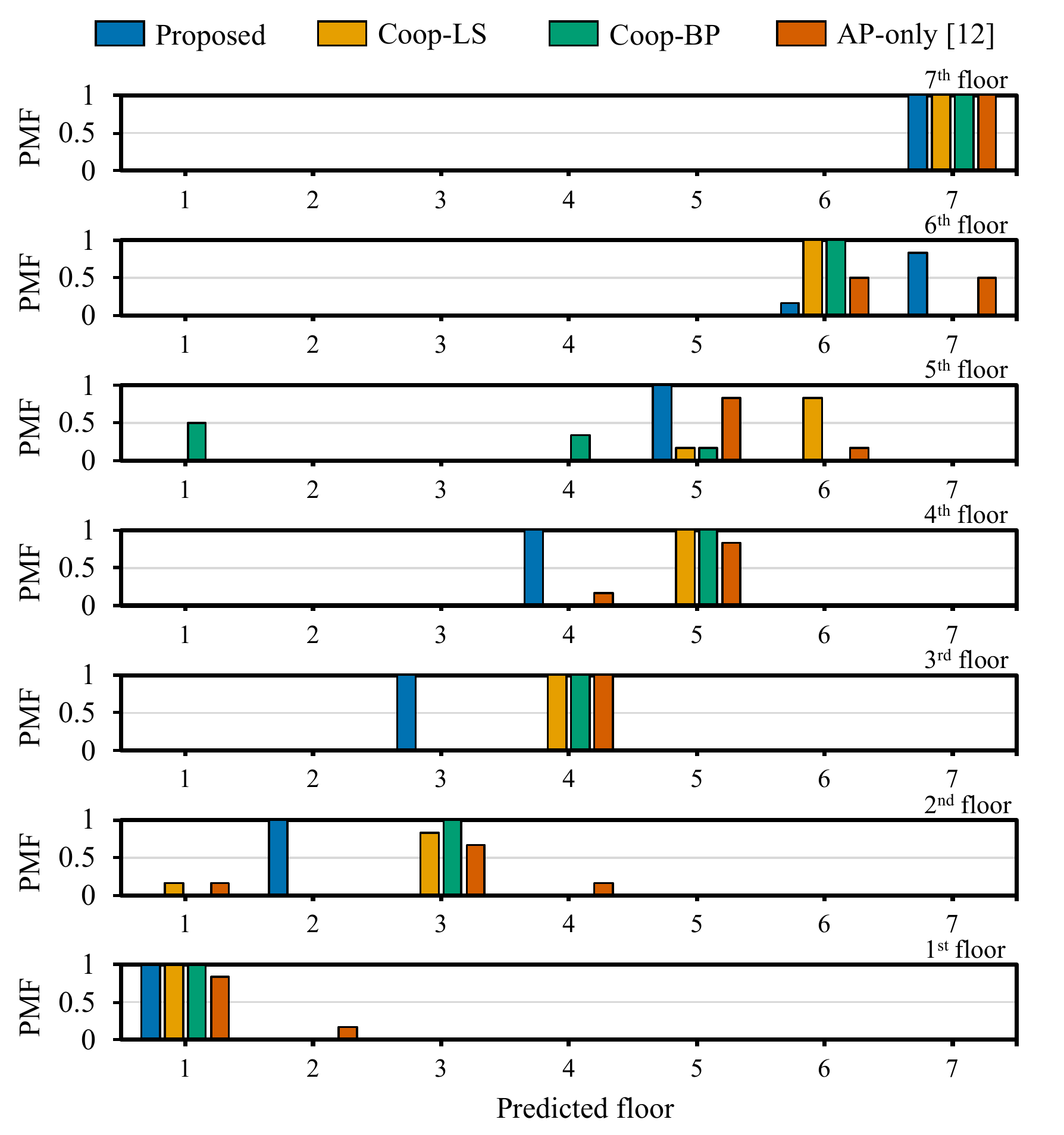}
        \label{fig:west2_pmf_ap11}
    }
    \caption{Predicted-floor PMFs in the realistic building
    model with $L=N-1$, using six fixed terminal positions
    per floor corresponding to those in the LoRa experiment:
    (a) AP4+4 and (b) AP1+1.
    Each row corresponds to a true floor, and the bars show
    the estimated-floor distribution for each method.}
    \label{fig:west2_pmf}
\end{figure}

The PMFs for the realistic building model in
Fig.~\ref{fig:west2_pmf} show more dispersed estimation errors than
those for the simplified model, reflecting the greater overlap among
floor-dependent RSS patterns in the more complex propagation
environment. The cooperative methods from the preliminary study remain
effective, but their error patterns vary more strongly with floor and
AP deployment. The proposed method concentrates more probability around
the correct floors by combining the proposed AP-side score with
statistical floor-difference information from the terminal links.

\begin{figure}[t]
    \centering
    \subfloat[AP4+4]{
        \includegraphics[width=0.9\linewidth]{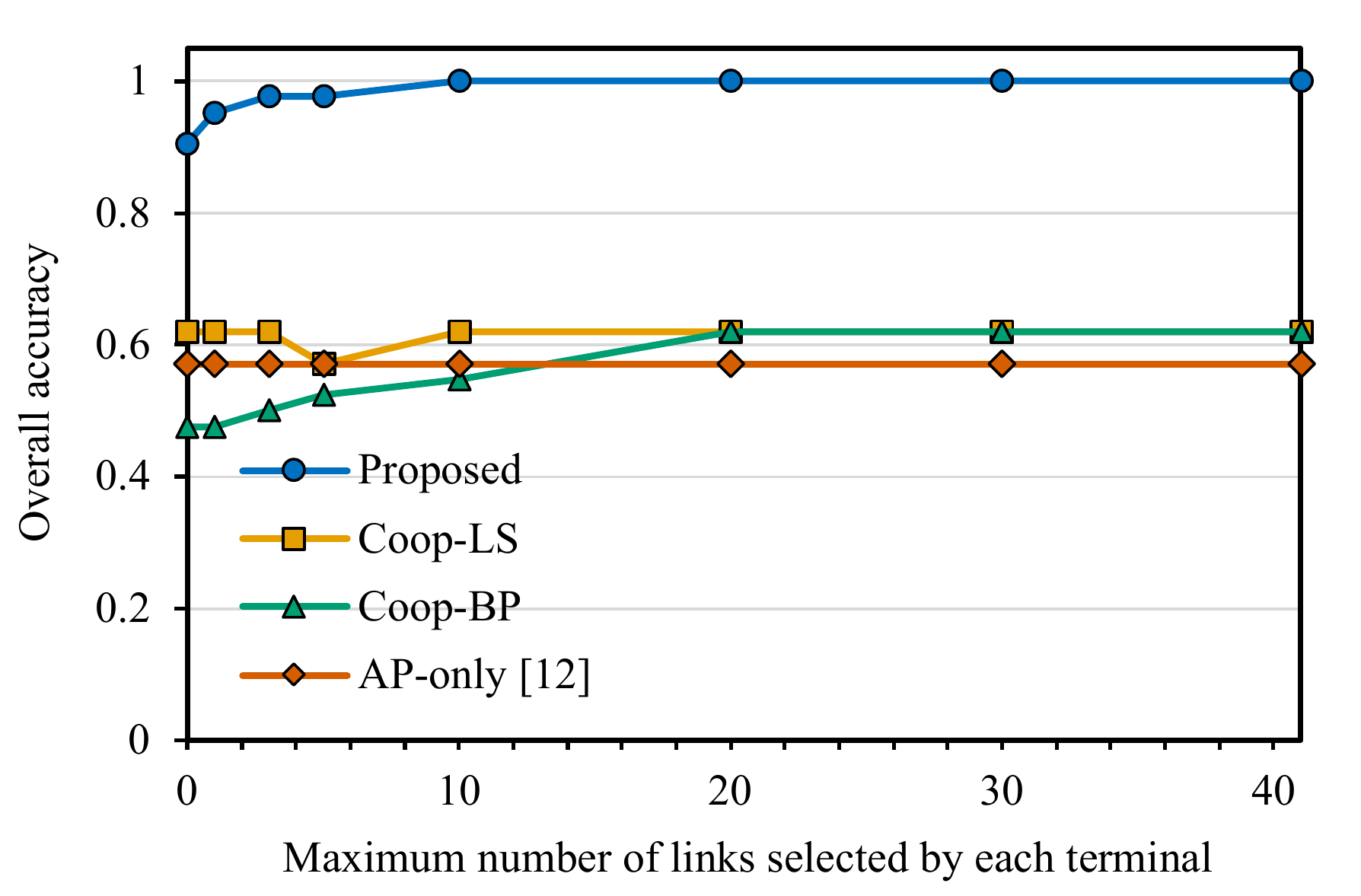}
        \label{fig:west2_links_ap44}
    }
    \vspace{-3mm}
    \subfloat[AP1+1]{
        \includegraphics[width=0.9\linewidth]{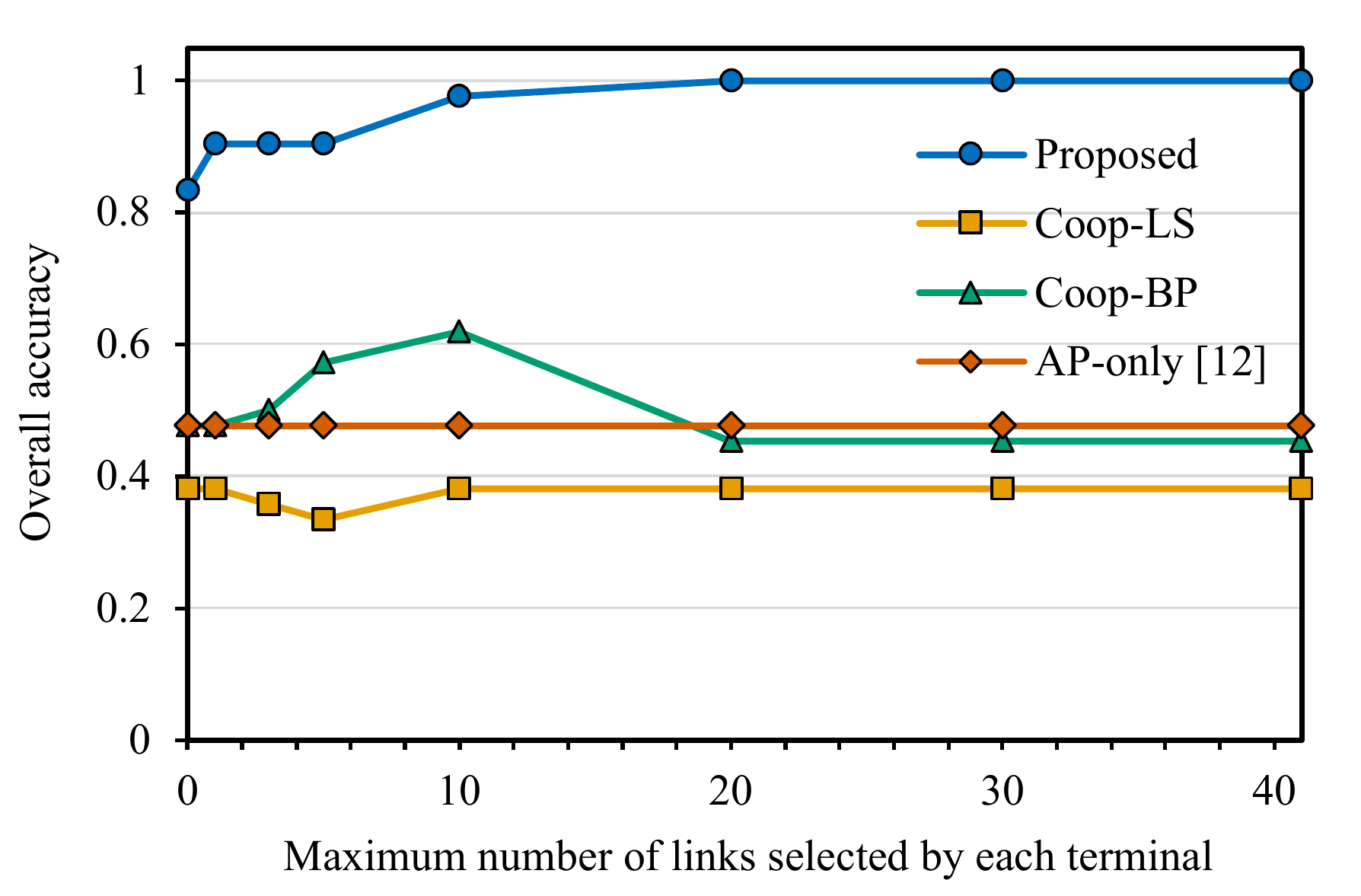}
        \label{fig:west2_links_ap11}
    }
    \caption{Overall floor-estimation accuracy versus
    terminal-selection limit $L$ in the realistic building
    model, using six fixed terminal positions per floor
    corresponding to those in the LoRa experiment:
    (a) AP4+4 and (b) AP1+1.
    AP4+4 uses four APs on each of floors 1 and 7;
    AP1+1 uses one AP on each of these floors.}
    \label{fig:west2_links}
\end{figure}

Fig.~\ref{fig:west2_links} further shows the effect of the
terminal-selection limit $L$ under different AP deployments.
Under AP4+4, the proposed method maintains higher accuracy
at smaller terminal-selection limits than under AP1+1.
With AP1+1, reducing $L$ has a larger effect because the AP-side
observations provide less floor information. These comparisons
refer to the realistic ray-tracing model.

\section{Experimental Evaluation}
\label{sec:experimental_evaluation}

\subsection{Experimental Setup}

The experiment was conducted in a seven-floor corridor environment. The
corridor is approximately 30\,m long and 2\,m wide, and the floor height
is 2.6\,m. 
Measurements were collected at six designated terminal
positions on each floor, resulting in 42 terminal positions in total.

A supplementary RSS survey was conducted at 16 positions per floor
with an approximately 2-m spacing. This survey is used only to
characterize the floor-wise distribution of the upper--lower AP RSS
contrast over the corridor and to examine whether 
the AP-side model estimated from the six terminal positions 
reproduces the broader environmental RSS characteristics. 
The supplementary survey samples are not used for model-parameter estimation 
or floor-estimation performance evaluation.

\begin{figure}[t]
\centering
\includegraphics[width=0.95\linewidth]{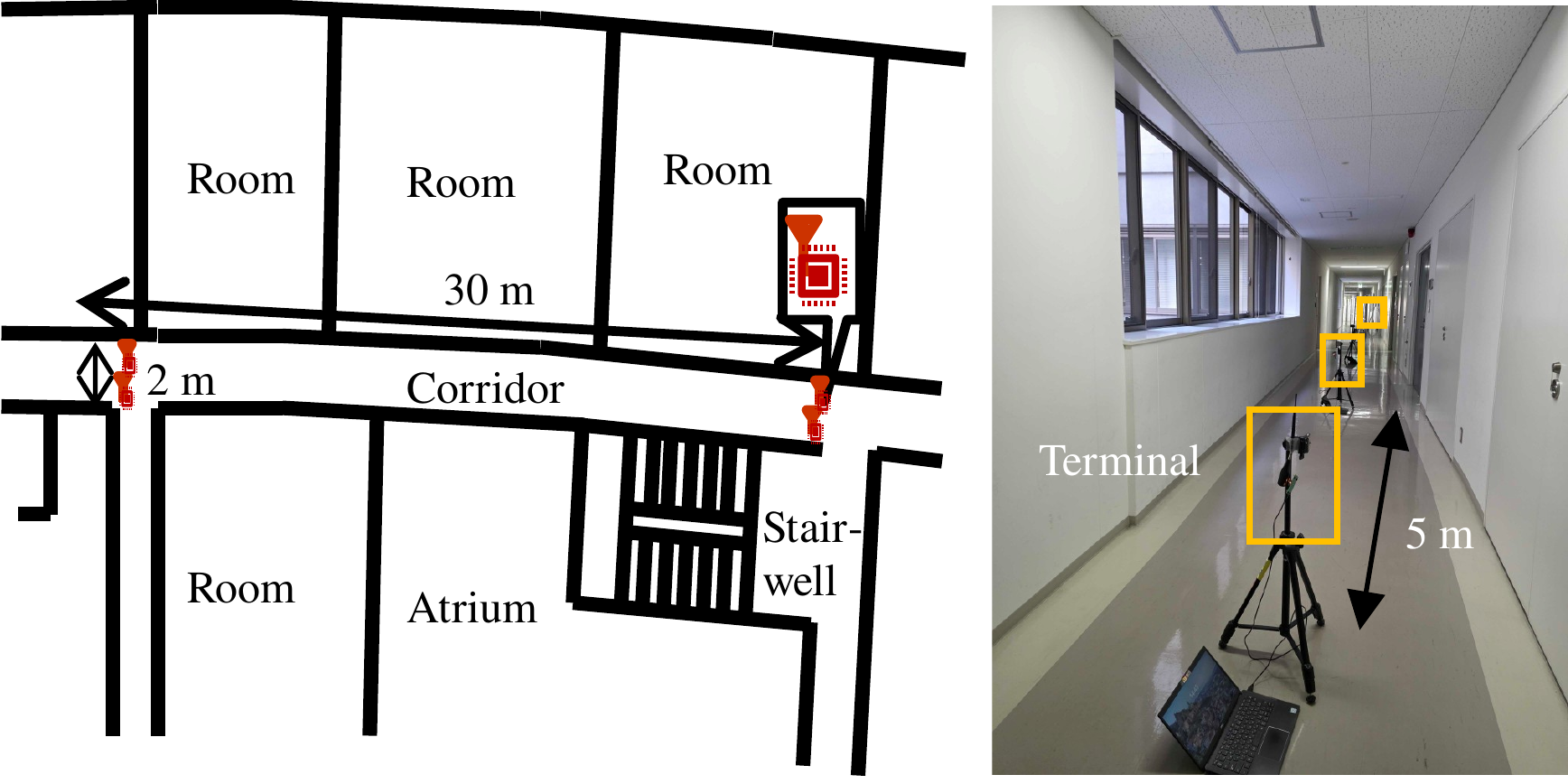}
\caption{Experimental environment and setup, including the floor plan of 
the corridor and a representative photograph showing the terminal placement.}
\label{fig:experimental_setup}
\end{figure}

The experimental environment and setup are shown in
Fig.~\ref{fig:experimental_setup}, including the floor plan of the
corridor and a representative photograph of the terminal placement.
Four APs are deployed on each of the first and seventh floors under
AP4+4, whereas one AP is used on each of these floors under AP1+1.
The LoRa experiment uses the same dual-order sequential ICM
procedure as the simulations, with at most three sweeps per order.
The AP-side and pairwise variance lower bounds are
$\sigma_{\mathrm{A,min}}^2=1\,\mathrm{dB}^2$ and
$\sigma_{\mathrm{P,min}}^2=9\,\mathrm{dB}^2$, respectively.

The radios operate at 920.7\,MHz with a 250-kHz bandwidth, spreading factor of 7, 
and transmit power of 13\,dBm. For this experimental evaluation,
AP-to-terminal and terminal-to-terminal RSS values are used in dBm,
while the upper--lower AP RSS contrast is expressed in dB.
AP packets are transmitted
at 1-s intervals, and RSS is collected for 20\,s at each measurement
position. Terminal-to-terminal RSS is also collected at the 42 terminal
positions for cooperative evaluation. 
Within each bootstrap draw, the corresponding AP-to-terminal and terminal-to-terminal 
observations are sampled together as one complete RSS feature vector.
The RSS samples used to calibrate both the AP-side and inter-terminal likelihood 
models were kept separate from those used for floor-level estimation.

For the experimental evaluation, let $b\in\{1,\ldots,B\}$
index the bootstrap trials, where $B=50$, and let
$\widehat h_i^{(b)}\in\mathcal K$ and $h_i^{\star}\in\mathcal K$
denote the estimated label in trial $b$ and the true floor label
at terminal position $i$, respectively. The absolute floor error is
$|\widehat h_i^{(b)}-h_i^{\star}|$. Its empirical cumulative
distribution function (CDF) represents the fraction of position--trial
estimates whose floor error does not exceed a given number of floors.

\begin{table}[t]
\caption{Experimental Parameters}
\label{tab:experiment}
\centering
\footnotesize
\renewcommand{\arraystretch}{1.08}
\begin{tabular}{@{}ll@{}}
\toprule
Parameter & Value \\
\midrule
Carrier frequency & 920.7 MHz \\
PHY & BW 250 kHz, SF7, 13 dBm \\
Building & 7 floors, 30 m $\times$ 2 m \\
Floor height & 2.6 m \\
Terminal positions & 6 per floor, 42 total \\
AP deployments & AP4+4 and AP1+1 \\
Collection & 20 s per position \\
AP interval & 1 s \\
Bootstrap trials & 50 \\
\bottomrule
\end{tabular}
\end{table}

\subsection{Validation of the AP-Side RSS Model}
\label{sec:calibration_validation}

\begin{figure}[t]
\centering
\includegraphics[width=0.99\linewidth]{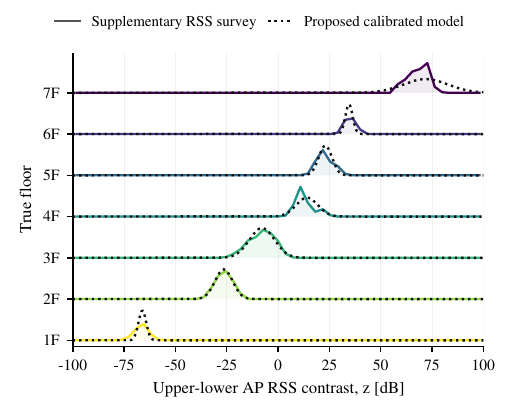}
\caption{Floor-wise distributions of the upper--lower AP RSS contrast $z$ under AP4+4: empirical distributions from the supplementary survey (16 positions per floor) and Gaussian models calibrated using six terminal positions per floor.}
\label{fig:z_model_validation}
\end{figure}

A supplementary RSS survey at 16 positions per floor was used to check whether 
the AP-side model estimated from the six terminal positions
captures the RSS variation over the corridor. In~\cite{li_vtc2026}, 
the RSS thresholds between adjacent floors
were set approximately by inspecting the measured RSS CDFs. 
In the present method, the AP-side score is instead constructed from 
a statistical model of the upper--lower AP RSS contrast estimated 
using the six terminal positions per floor. 
Under AP4+4, Fig.~\ref{fig:z_model_validation}
compares the resulting distributions of the upper--lower AP RSS contrast
$z$ with those from the supplementary survey.

The fitted distributions follow the main peaks and the floor-dependent
shift observed in the supplementary survey. Some local variations are
smoothed, especially where the measured RSS is multimodal, but the
overall trend is well preserved.
We also examined inter-terminal RSS grouped by floor difference.
The Gaussian models reasonably capture the main distribution trends
and dispersion, although deviations remain; the corresponding plots
are omitted for brevity.

\subsection{Comparison with Path-Loss-Based Methods}
\begin{figure}[t]
\centering
\includegraphics[width=0.85\linewidth]{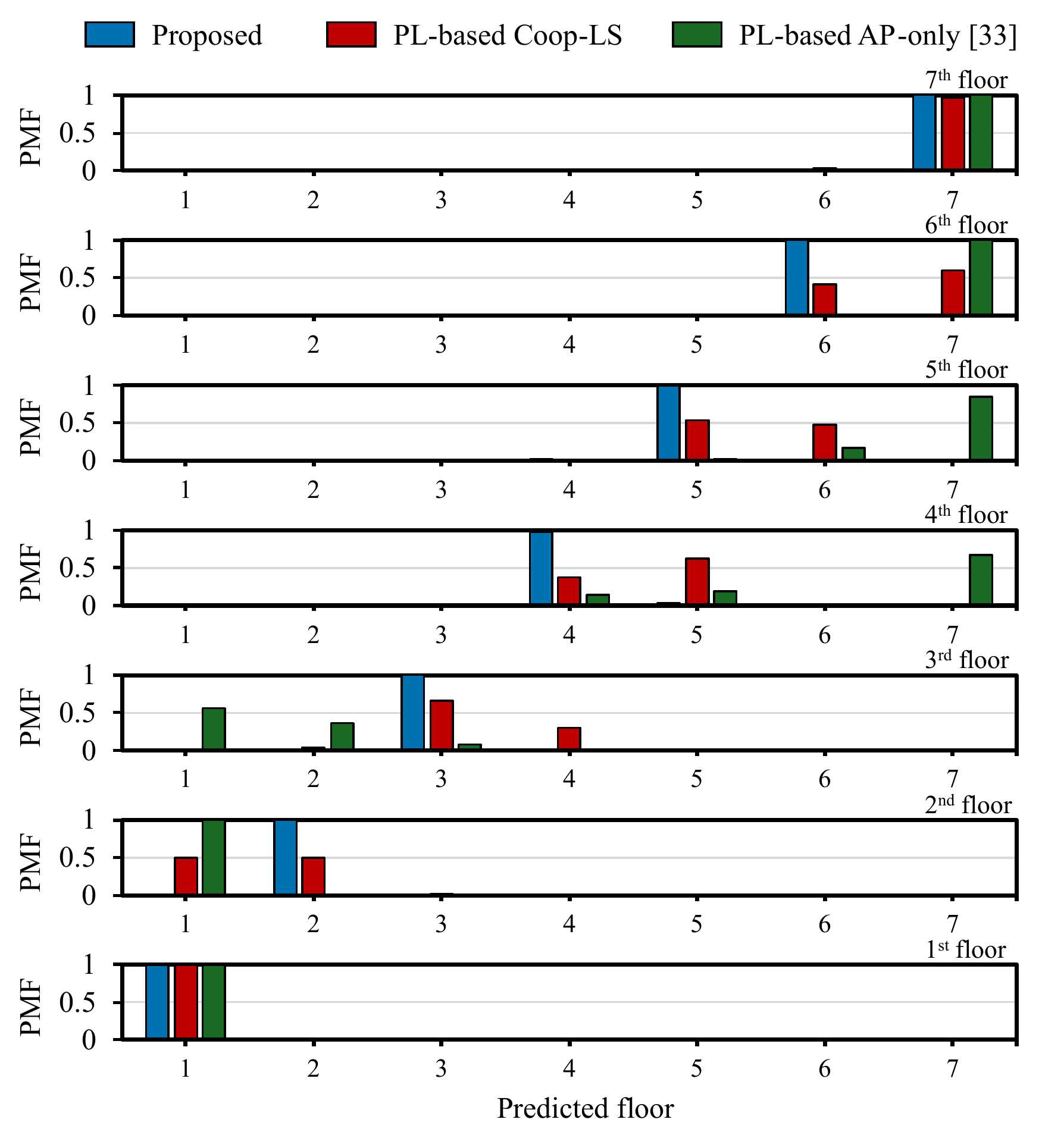}
\caption{Predicted-floor PMFs of the proposed and path-loss-based
methods under AP4+4 with $L=N-1$. PL-based AP-only uses
the model in~\cite{vo2024_pathloss}, and PL-based Coop-LS is its
cooperative extension considered here.}
\label{fig:pathloss_comparison}
\end{figure}

Fig.~\ref{fig:pathloss_comparison} compares the proposed method with PL-based 
AP-only and PL-based Coop-LS under AP4+4. The path-loss model supplies a physically 
interpretable RSS-to-distance mapping, but a single monotonic relationship between 
RSS and distance is difficult to maintain in the presence of multipath propagation, 
floor penetration, and wall-reflected paths. Adding terminal cooperation through 
PL-based Coop-LS improves some floor decisions, demonstrating that the path-loss 
observations can benefit from relative constraints. The proposed method produces 
more concentrated PMFs because it directly models floor-dependent RSS as statistical 
evidence rather than converting RSS into geometric distance estimates.

\subsection{Bootstrap Floor-Wise Accuracy and Error CDF}
\begin{figure}[t]
    \centering

    \subfloat[Floor-wise estimation accuracy (AP4+4)]{
        \includegraphics[width=0.95\linewidth]{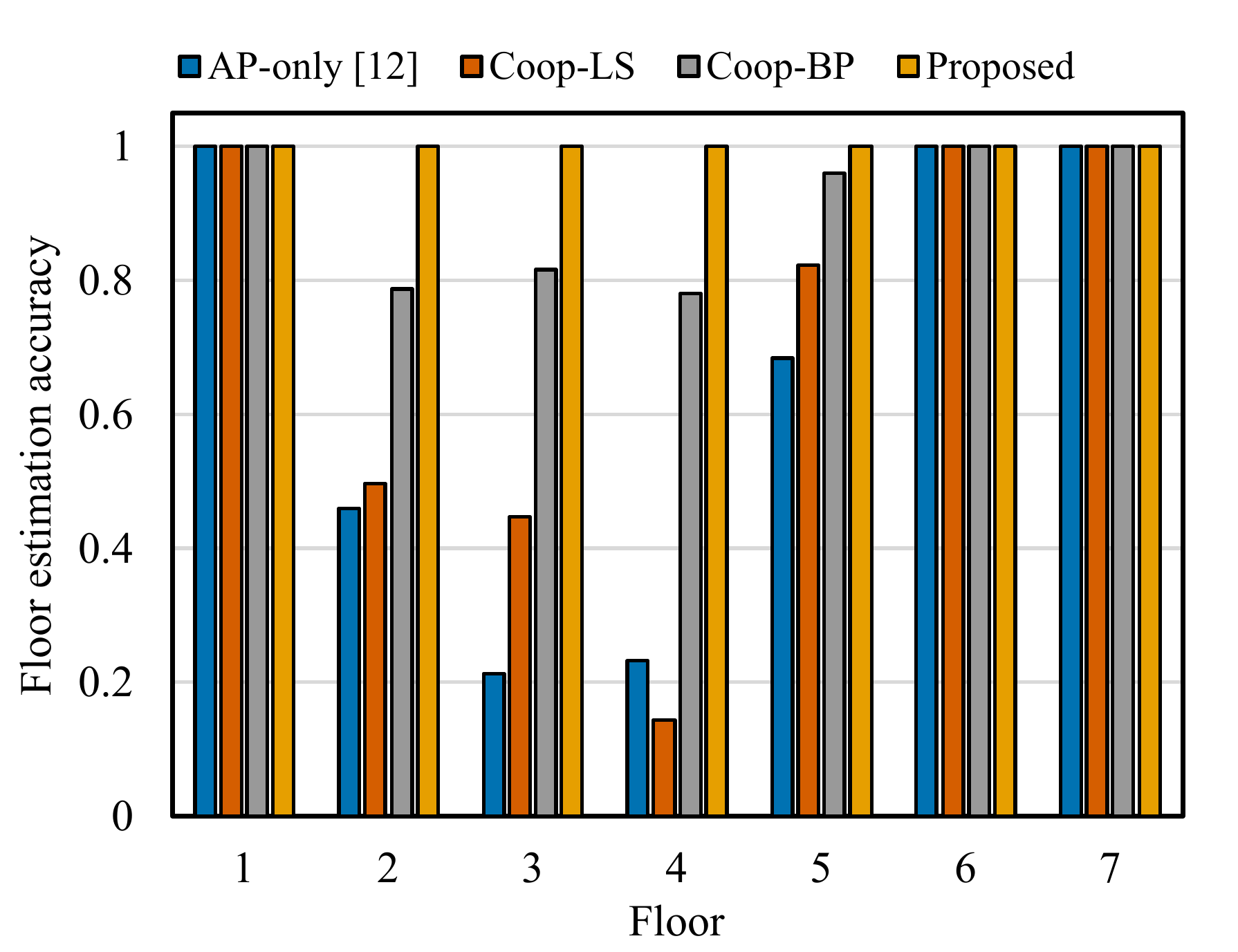}
        \label{fig:experimental_summary_accuracy}
    }

    \vspace{1mm}

    \subfloat[Error CDF (AP1+1)]{
        \includegraphics[width=0.47\linewidth]{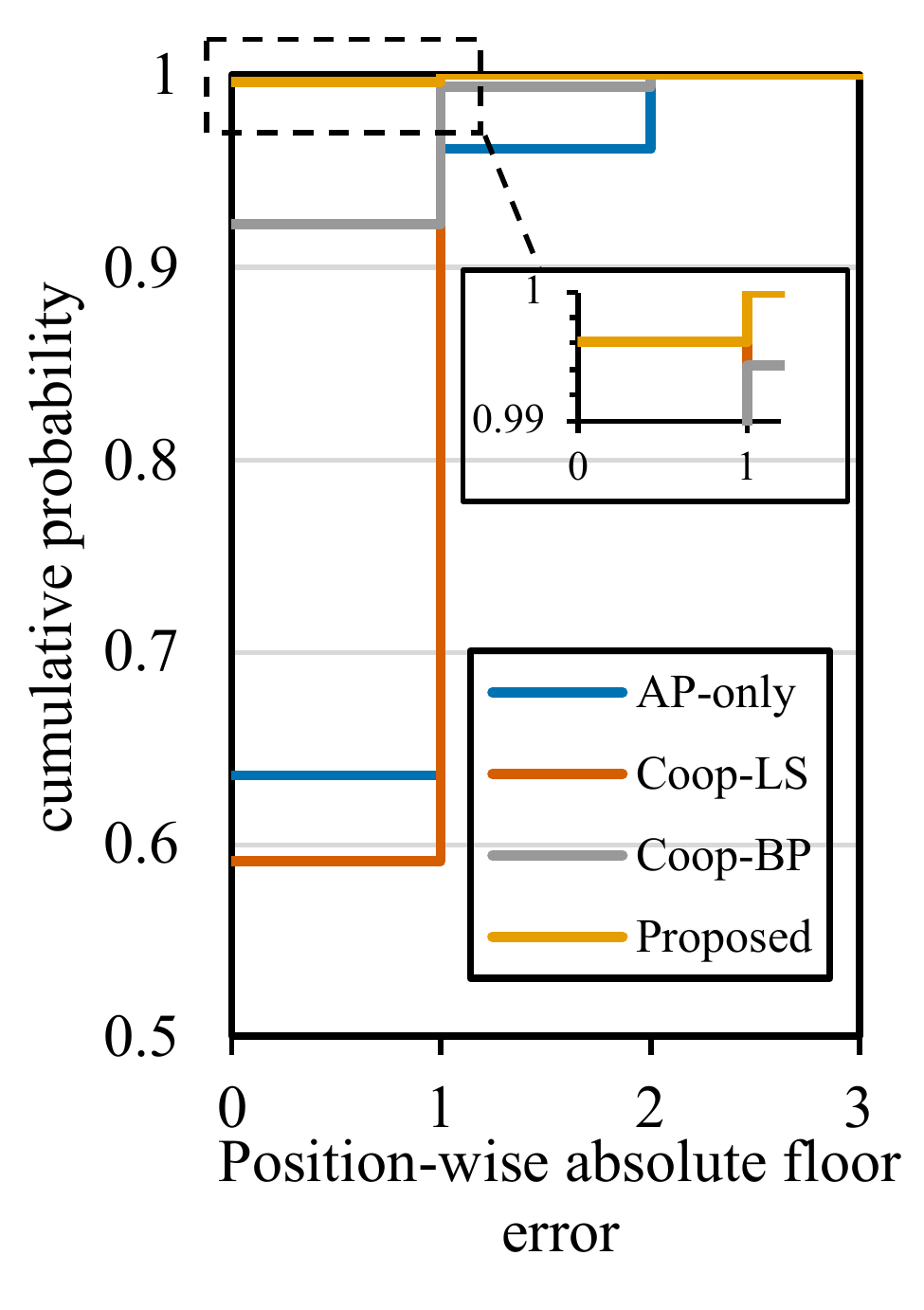}
        \label{fig:experimental_summary_cdf_ap11}
    }
    \hfill
    \subfloat[Error CDF (AP4+4)]{
        \includegraphics[width=0.47\linewidth]{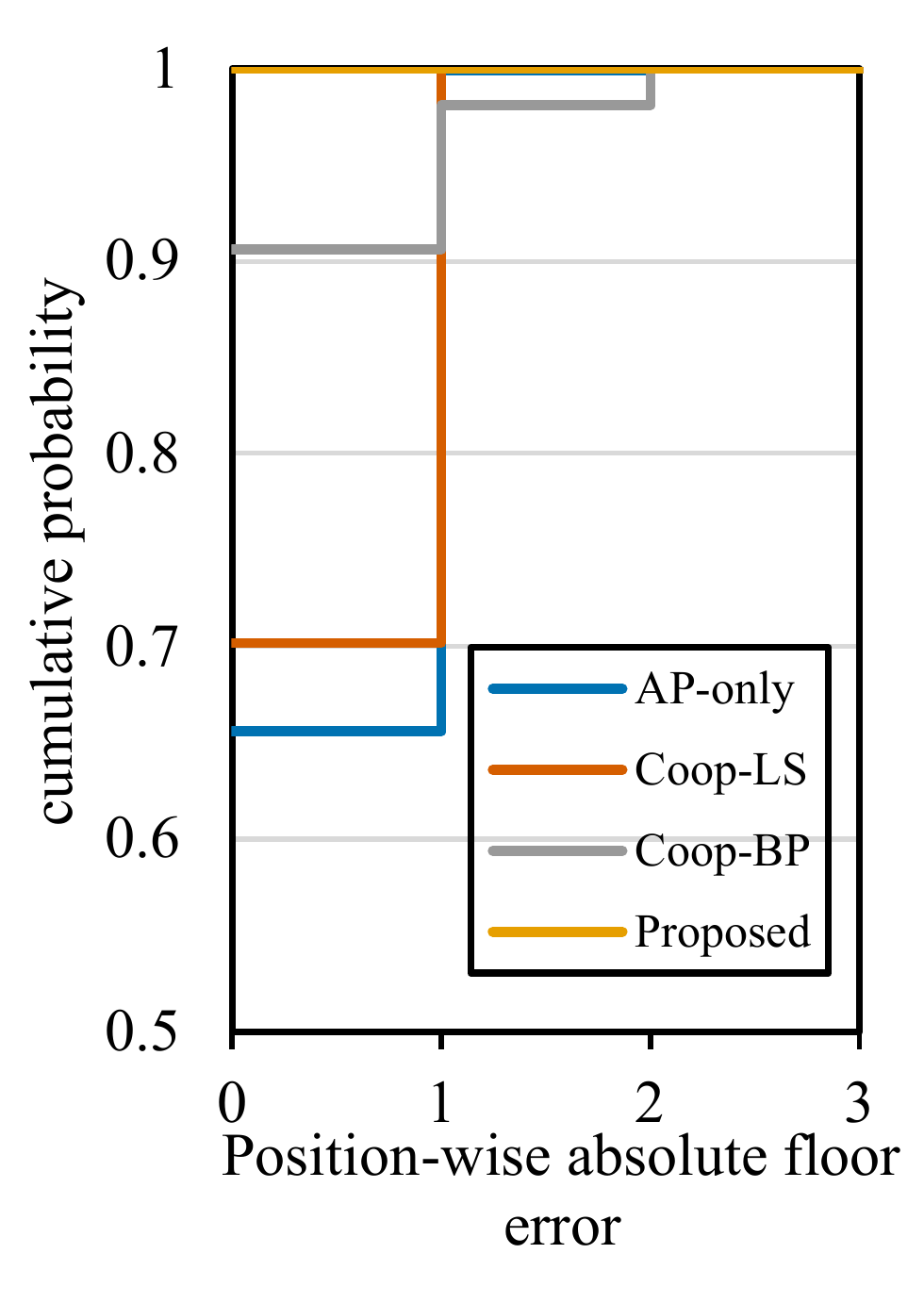}
        \label{fig:experimental_summary_cdf_ap44}
    }

    \caption{Experimental floor-estimation results with
    $L=N-1$ over 50 bootstrap trials:
    (a) mean floor-wise estimation accuracy under AP4+4;
    (b) empirical CDF of absolute floor error under AP1+1;
    and (c) empirical CDF of absolute floor error under AP4+4.}
    \label{fig:experimental_summary}
\end{figure}

Fig.~\ref{fig:experimental_summary}(a) reports mean floor-wise
estimation accuracy under AP4+4. 
AP-only remains accurate on floors with distinctive AP-side 
RSS but degrades on floors where upper- and lower-AP evidence overlaps. 
The cooperative methods in our preliminary study improve several of these floors, 
confirming the value of terminal-to-terminal observations. The proposed method 
achieves stable floor-wise accuracy by combining the AP-side floor score with 
the cooperative floor-difference score.

Figs.~\ref{fig:experimental_summary}(b) and (c) show the
empirical cumulative distributions of absolute floor error
under AP1+1 and AP4+4, respectively. The inset in (b) highlights 
differences in the cumulative probability near one. 
The CDF at an integer threshold
represents the fraction of position--trial estimates whose
absolute floor error does not exceed that threshold.
Under AP4+4, the proposed method achieves zero
floor-estimation error over the evaluated samples.
Under AP1+1, residual errors become more apparent,
allowing the methods to be compared under sparser AP deployment.

\subsection{Experimental PMFs and Terminal Selection}
\begin{figure}[t]
    \centering
    \subfloat[AP4+4]{
        \includegraphics[width=0.85\linewidth]{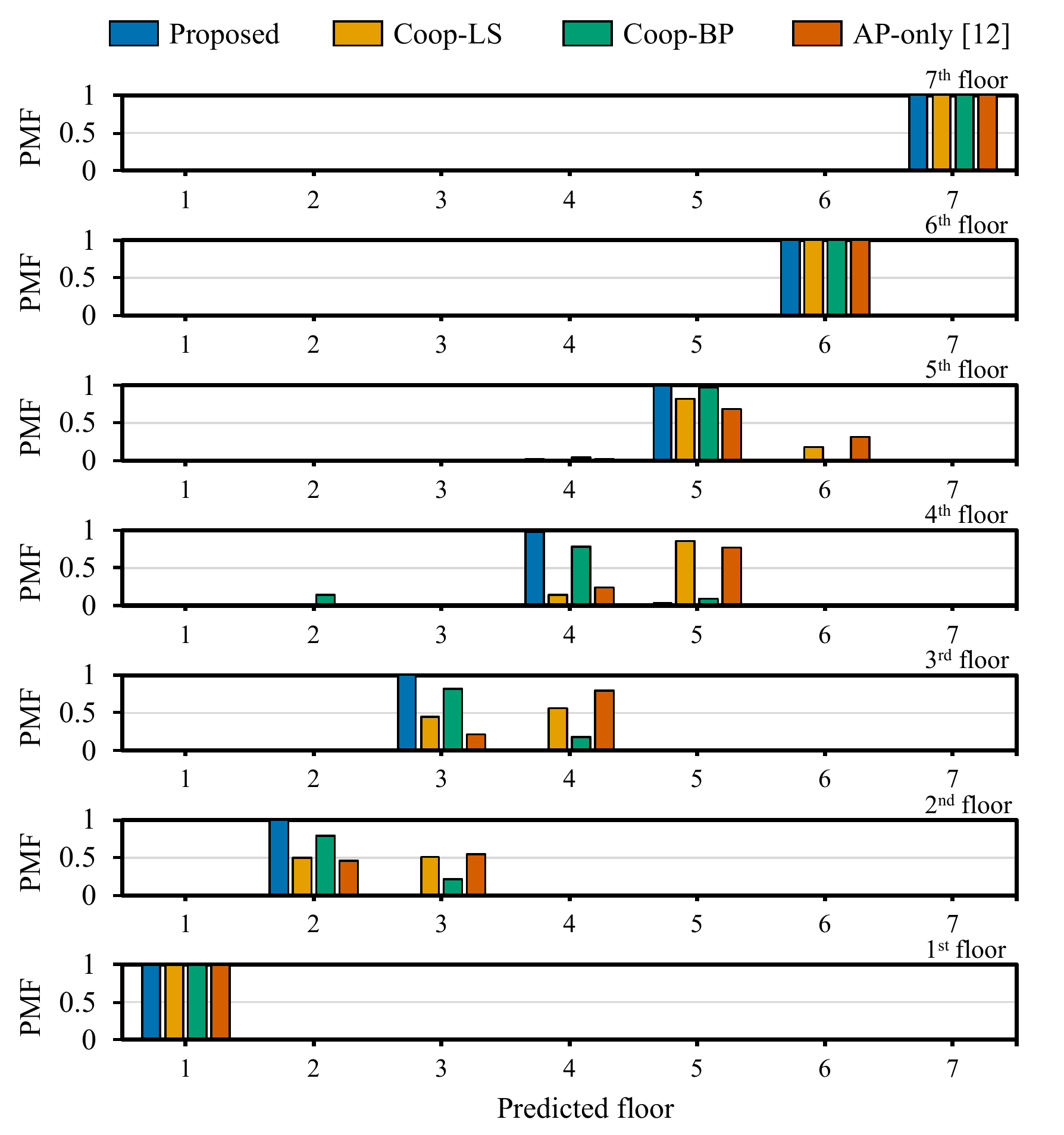}
        \label{fig:experimental_pmf_ap44}
    }
    \vspace{-3mm}
    \subfloat[AP1+1]{
        \includegraphics[width=0.85\linewidth]{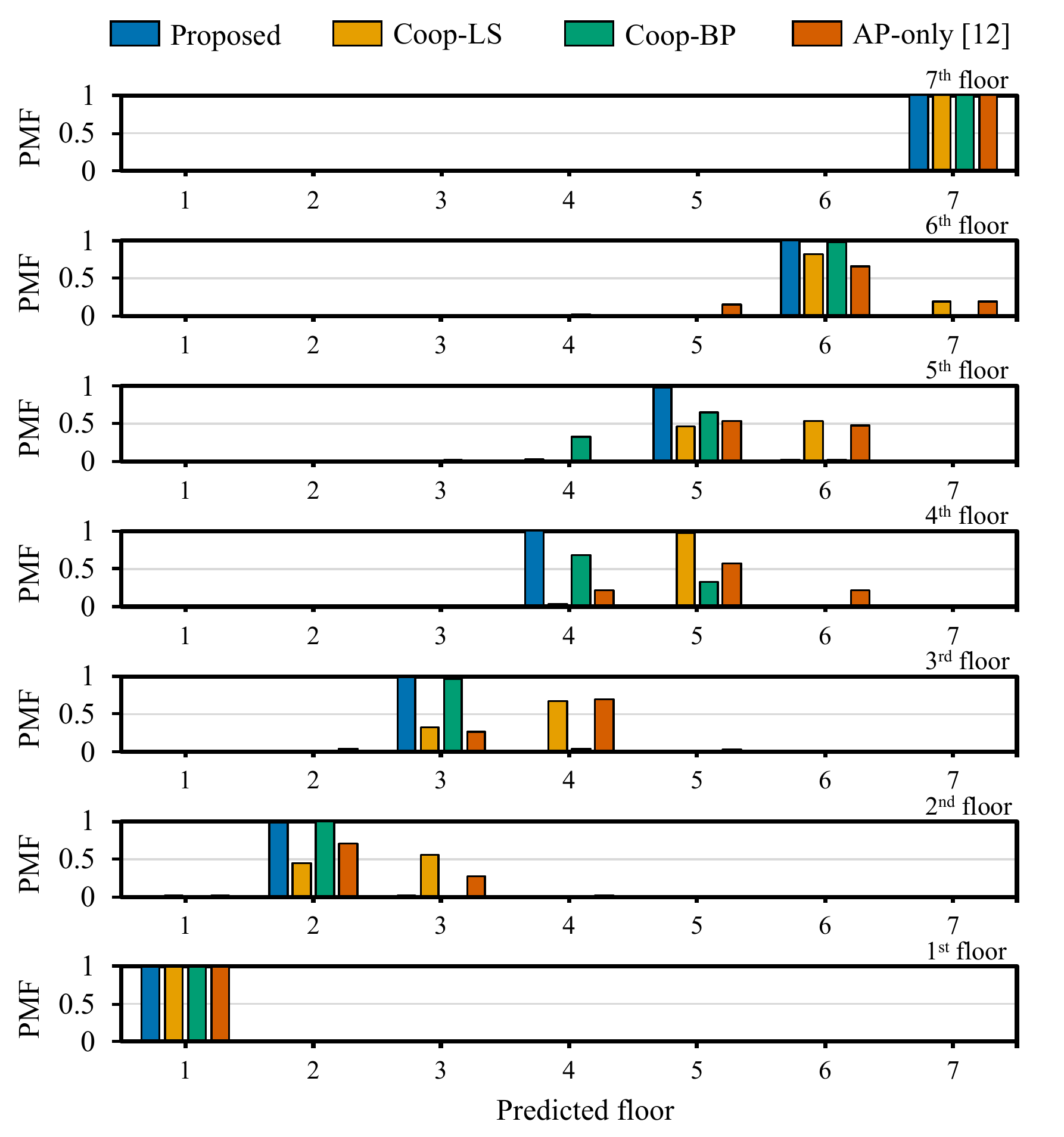}
        \label{fig:experimental_pmf_ap11}
    }
    \caption{Predicted-floor PMFs on the experimental dataset with
    $L=N-1$ under (a) AP4+4 and (b) AP1+1. Each row
    corresponds to a true floor, and the bars show estimated-floor
    distributions over repeated RSS samples.}
    \label{fig:experimental_pmf}
\end{figure}

The predicted-floor PMFs in Fig.~\ref{fig:experimental_pmf}
show the different effects of AP deployment on the methods.
Under AP4+4, the AP-only and preliminary
cooperative methods exhibit errors mainly on intermediate floors,
where AP-side RSS patterns overlap, whereas the proposed method
correctly classifies all evaluated terminal positions.
Under AP1+1, the sparser AP deployment increases
floor ambiguity; nevertheless, the proposed method
assigns a larger empirical proportion of estimates
to the correct floor than the AP-only method and 
the preliminary cooperative methods in~\cite{li_vtc2026}.

\begin{figure}[t]
    \centering
    \subfloat[AP4+4]{
        \includegraphics[width=0.9\linewidth]{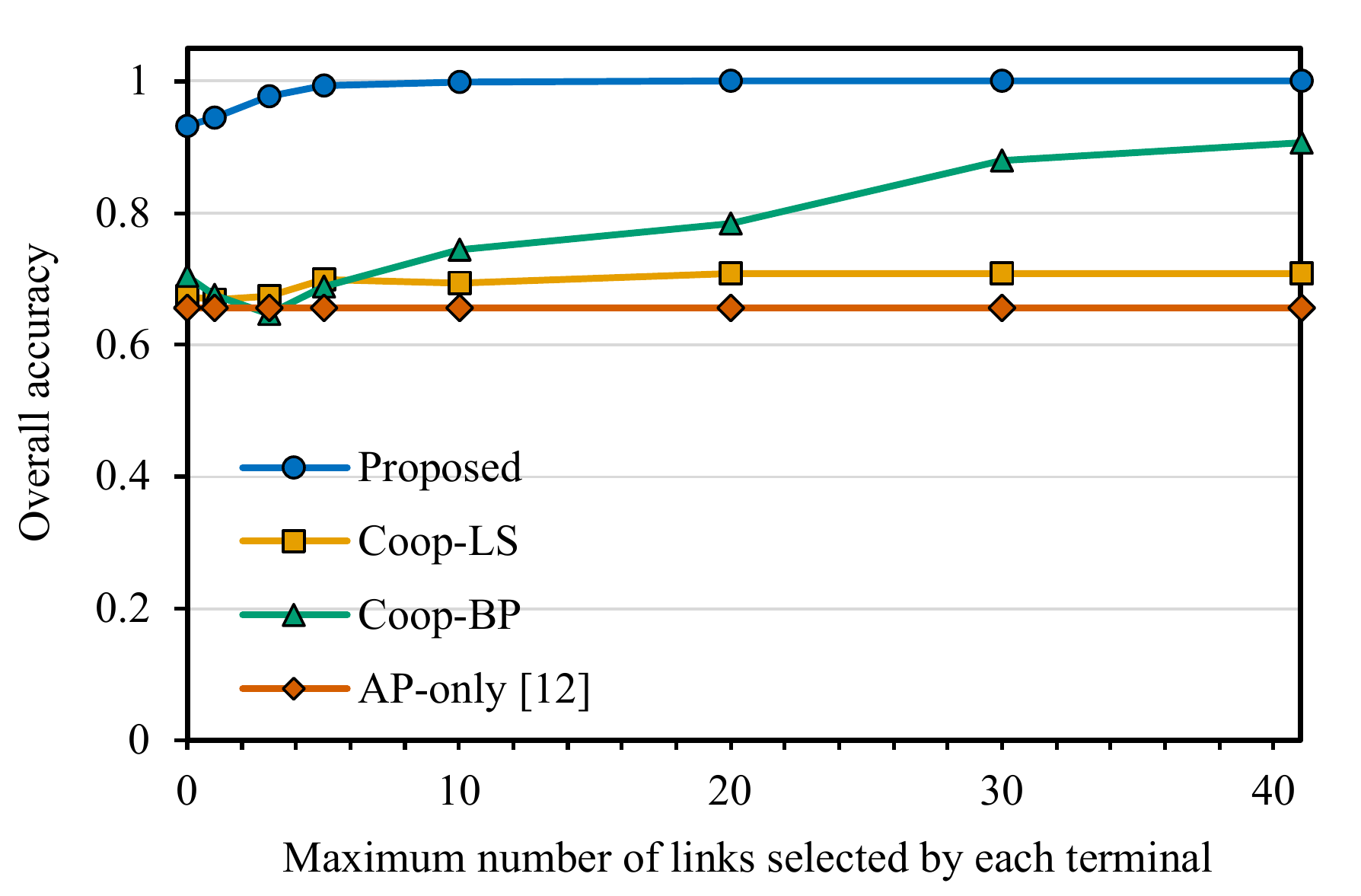}
        \label{fig:experimental_links_ap44}
    }
    \vspace{-3mm}
    \subfloat[AP1+1]{
        \includegraphics[width=0.9\linewidth]{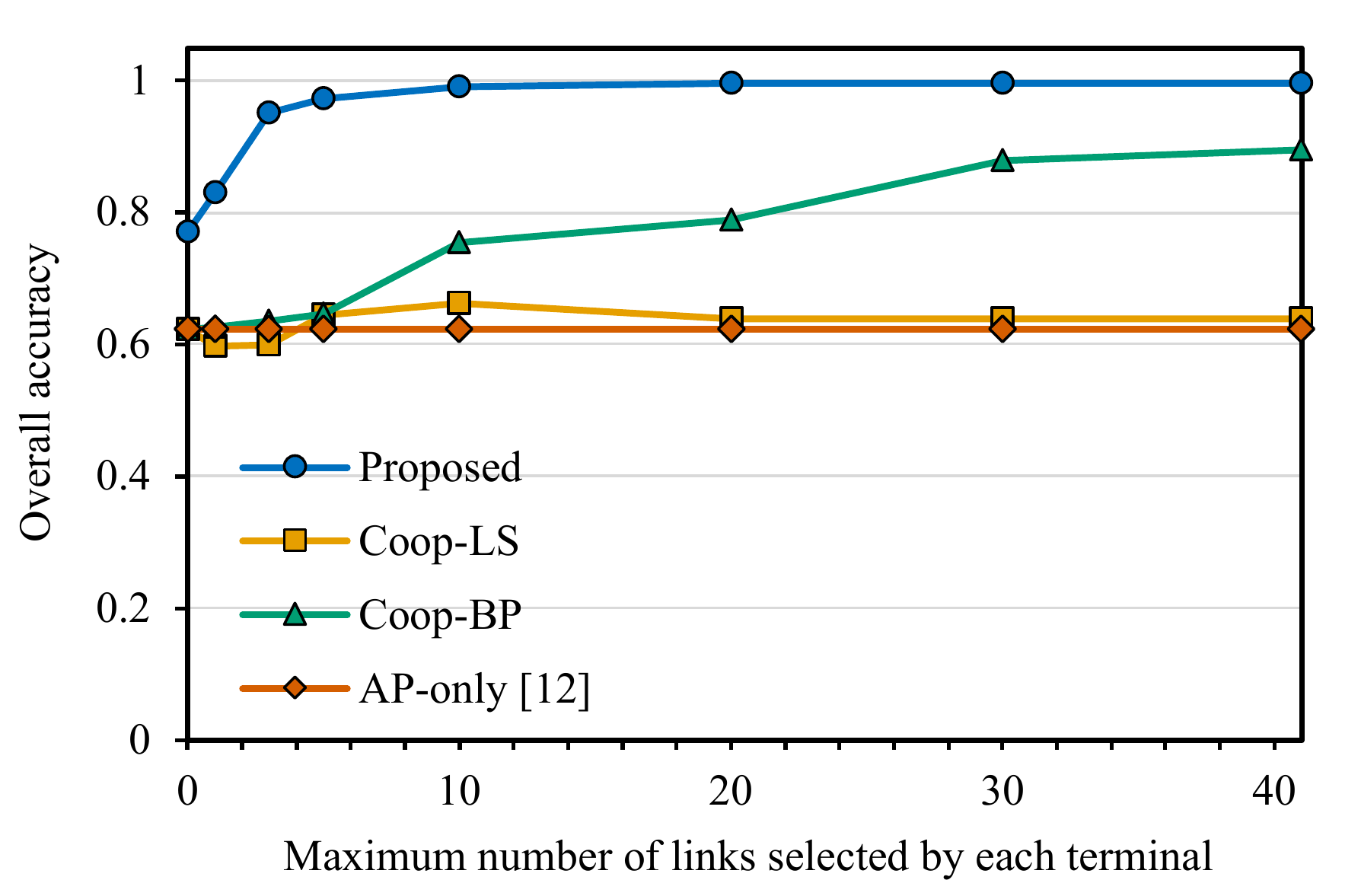}
        \label{fig:experimental_links_ap11}
    }
    \caption{Overall floor-estimation accuracy versus terminal-selection limit $L$ in the LoRa experiment: (a) AP4+4 and (b) AP1+1. AP4+4 uses four APs on each of floors 1 and 7; AP1+1 uses one AP on each. Observation-model parameters remain fixed as $L$ varies.}
    \label{fig:experimental_links}
\end{figure}

Fig.~\ref{fig:experimental_links} shows the effect of the
terminal-selection limit $L$ with fixed observation models.
At $L=0$, no inter-terminal observations are used, but the methods
retain their respective AP-side estimators. The proposed method reduces to its calibrated AP-side estimator,
whereas the baseline methods retain their respective AP-side
formulations; their $L=0$ accuracies therefore differ. Under
AP1+1, the proposed method improves accuracy from 77.10\%
at $L=0$ to 97.19\% at $L=5$.

In the LoRa measurements, 100\%, 99.65\%, 96.35\%, and 49.86\%
of the \emph{directed terminal selections} correspond to
same-floor terminal pairs at $L=1,3,5,10$, respectively. 
These percentages refer to selections before OR symmetrization, 
not to the degrees or edge proportions of the final undirected graph. 
With six terminals per floor, each terminal has at most
five same-floor candidates. The improvement under AP1+1
therefore indicates that predominantly same-floor constraints
can help resolve AP-side ambiguity.
Table~\ref{tab:controlled_link_mix} compares prescribed link
compositions at $L=5$, with accuracy averaged equally over the
two ray-tracing scenarios and the LoRa measurements. These
diagnostic mixtures use true floor labels, whereas the practical
link-selection rule uses only RSS.

\begin{table}[t]
\caption{Mean floor-estimation accuracy (\%) of the proposed
method by terminal-selection composition ($L=5$).}
\label{tab:controlled_link_mix}
\centering
\footnotesize
\renewcommand{\arraystretch}{1.08}
\setlength{\tabcolsep}{2.5pt}
\begin{tabular}{@{}l*{7}{c}@{}}
\toprule
 & \multicolumn{6}{c}{Same-floor:cross-floor terminal selections} & \\
\cmidrule(lr){2-7}
APs & 0:5 & 1:4 & 2:3 & 3:2 & 4:1 & 5:0 & RSS-ranked \\
\midrule
AP4+4 & 95.78 & 95.63 & \textbf{97.46} & 96.37 & 96.59 & 95.62 & 95.65 \\
AP1+1 & 90.78 & 90.17 & 91.54 & \textbf{93.51} & 92.44 & 91.87 & 91.84 \\
\bottomrule
\end{tabular}
\end{table}

Among the evaluated compositions, the highest mean accuracy
is 97.46\% for AP4+4 at 2:3 and 93.51\% for AP1+1 at 3:2.
The prescribed compositions use true floor labels for diagnostic
comparison; the practical RSS-ranked rule does not require them.

\subsection{Effect of the AP-Side Observation Model on Coop-BP}
\begin{figure}[t]
    \centering
    \subfloat[AP4+4]{
        \includegraphics[width=0.9\linewidth]{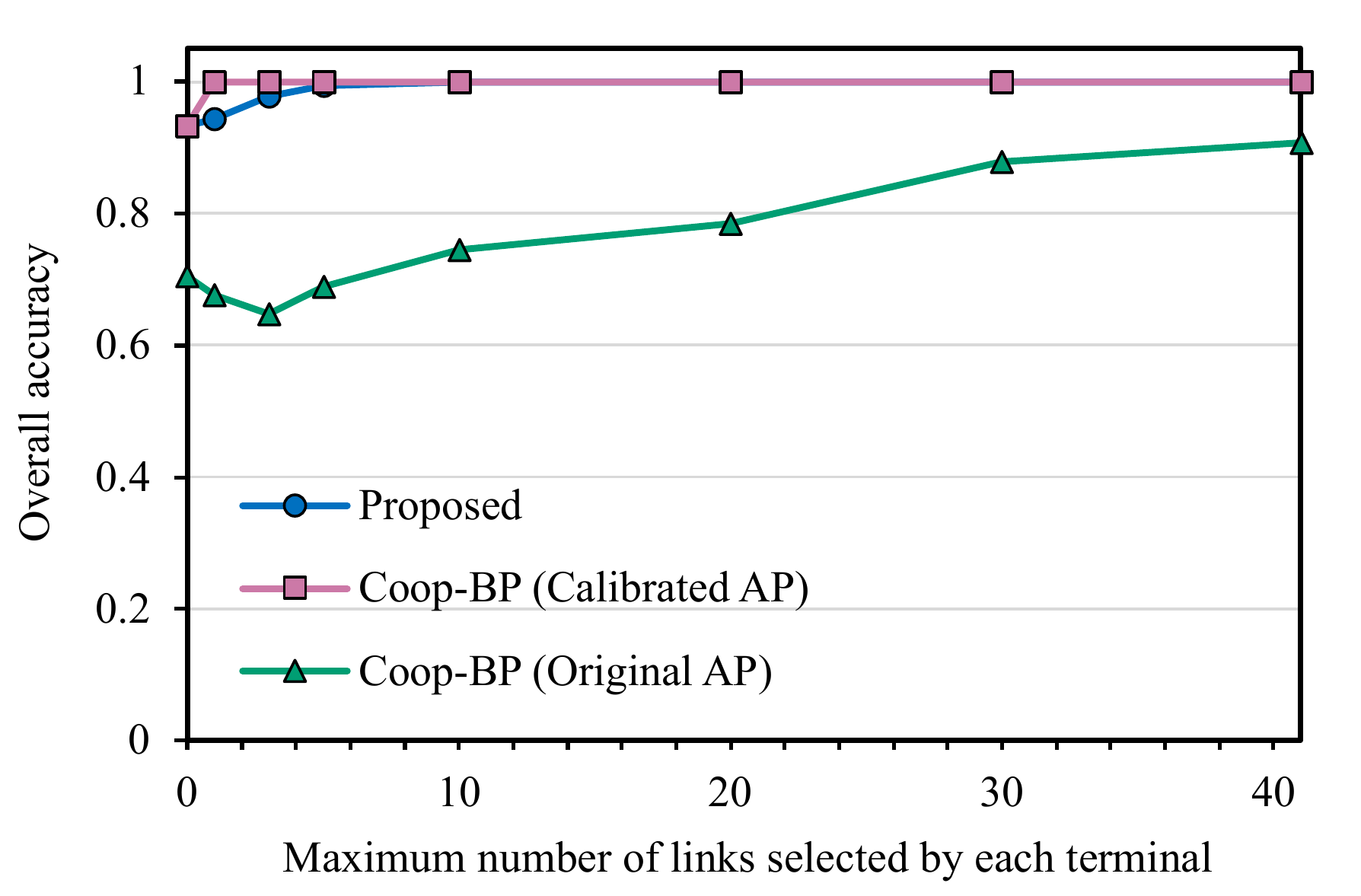}
        \label{fig:bp_calibration_ap44}
    }
    \vspace{-3mm}
    \subfloat[AP1+1]{
        \includegraphics[width=0.9\linewidth]{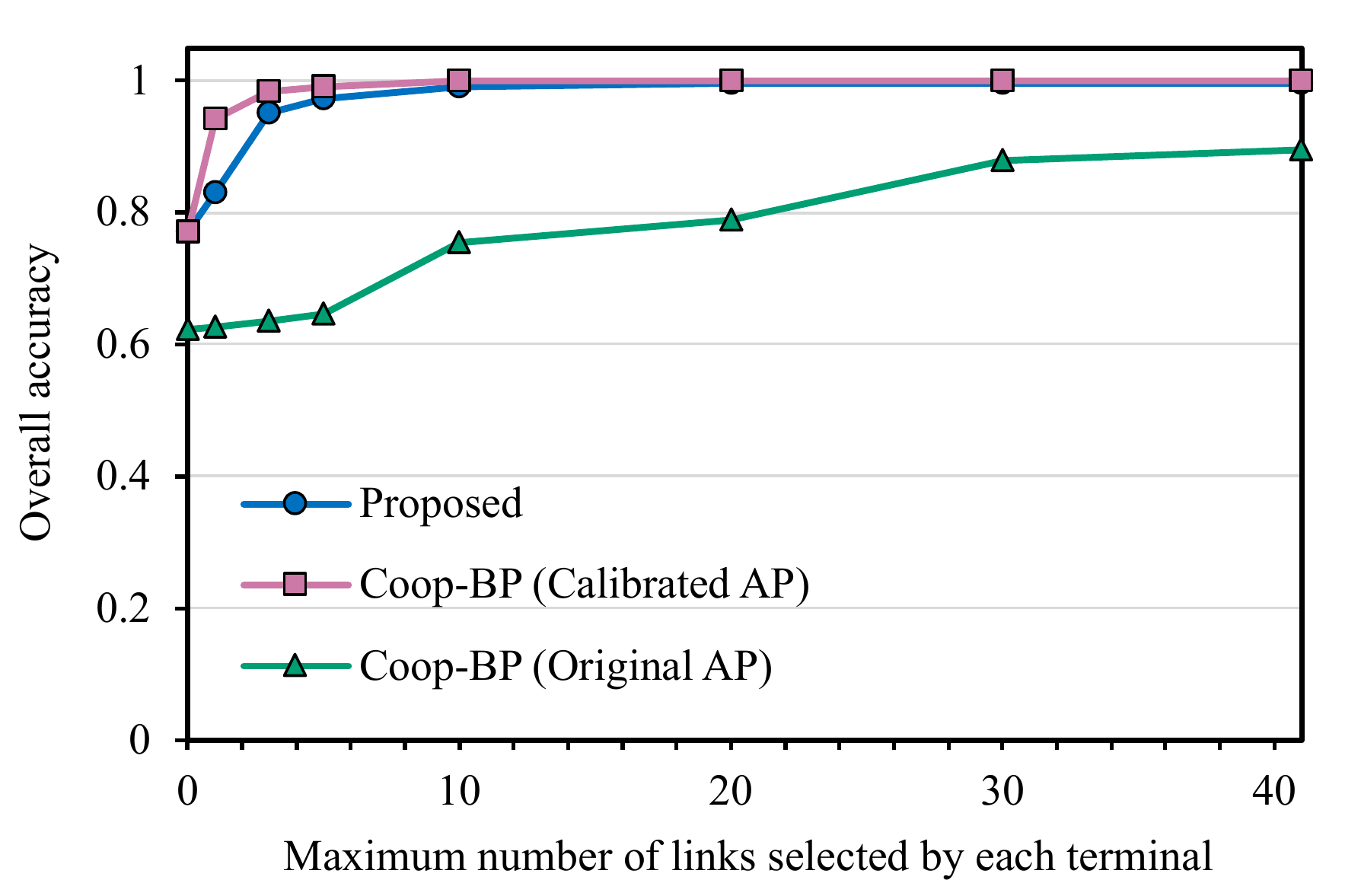}
        \label{fig:bp_calibration_ap11}
    }
    \caption{Effect of AP-side calibration on Coop-BP versus terminal-selection limit $L$ in the LoRa experiment: (a) AP4+4 and (b) AP1+1. AP4+4 uses four APs on each of floors 1 and 7; AP1+1 uses one AP on each. The two Coop-BP variants use the same BP inference rule and pairwise observation model.}
    \label{fig:bp_calibration}
\end{figure}

The two Coop-BP curves in Fig.~\ref{fig:bp_calibration}
compare the original and calibrated AP-side information while
retaining the same BP inference and pairwise model.
Coop-BP (Original AP) uses the original AP-side information
from our preliminary study~\cite{li_vtc2026}, whereas
Coop-BP (Calibrated AP) uses the proposed AP-side score.
Their accuracy difference, particularly under AP1+1 or at
larger $L$, demonstrates the importance of the AP-side
observation model in cooperative BP inference.

The complete proposed method additionally uses its own
pairwise statistical scores, constructs the OR-symmetrized
cooperation graph, and performs dual-order sequential ICM on
the weighted objective. Therefore, Fig.~\ref{fig:bp_calibration}
is an analysis of AP-side calibration within the preliminary
BP framework~\cite{li_vtc2026}, not an isolated comparison 
of the proposed ICM solver with BP.

\subsection{Computation Time}
\begin{figure}[t]
\centering
\includegraphics[width=0.9\linewidth]{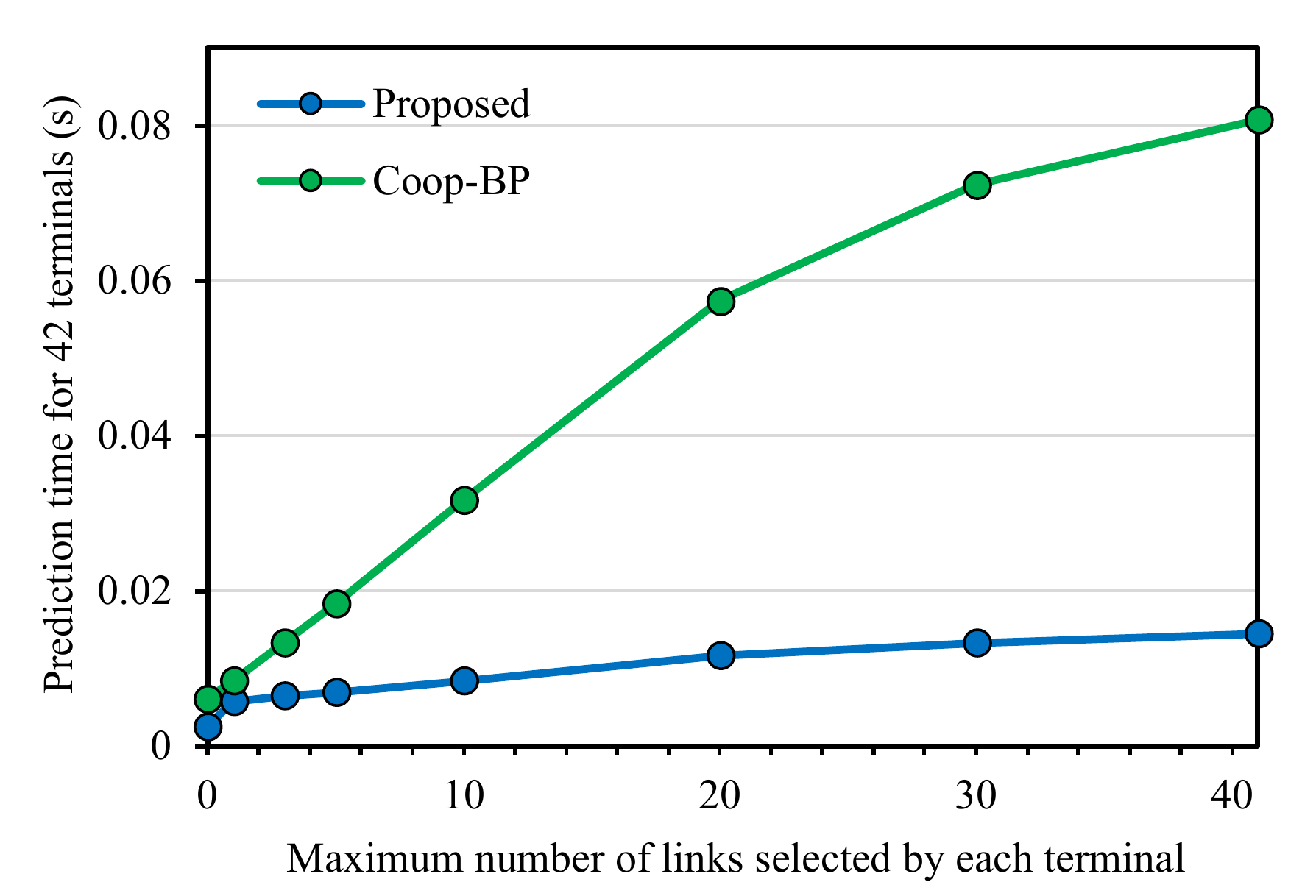}
\caption{Median runtime of the complete online prediction pipeline for 42 terminals versus terminal-selection limit $L$ under AP4+4 (30 runs on an Intel Core Ultra 5 225U processor). Coop-BP uses the original method from~\cite{li_vtc2026}.}
\label{fig:runtime}
\end{figure}

Fig.~\ref{fig:runtime} compares online prediction time for all
42 terminals under AP4+4, measured using an Intel Core Ultra 5
225U processor. The Coop-BP baseline uses the original method from
our preliminary study~\cite{li_vtc2026}, without the AP-side
calibration introduced in this paper. At $L=0$, the absolute time
difference between methods is relatively small, whereas the gap increases as the
terminal-selection limit $L$ grows. Coop-BP becomes more
expensive because its message updates evaluate pairs of floor
states, whereas the proposed ICM evaluates candidate labels
with the labels of connected terminals fixed.
At the maximum terminal-selection limit, the median prediction
times are approximately 0.0146~s for the proposed method and
0.0808~s for Coop-BP, making the latter about 5.53 times slower.
The timing includes graph-related processing and inference.

\subsection{Discussion}
The LoRa evaluation uses one seven-floor corridor and one carrier
frequency, with known AP floor labels and available inter-terminal
links. Applying the calibrated RSS models to another building requires
site-specific measurements.

\section{Conclusion}
This paper extended the device-centric cooperative floor-level
estimation framework by constructing statistical observation
models for AP-to-terminal and terminal-to-terminal RSS and
using dual-order sequential ICM to approximately optimize
the resulting weighted floor-label objective. Ray-tracing simulations and
LoRa measurements evaluated accuracy across AP deployments and
terminal-selection limits. In the AP1+1 LoRa experiment, accuracy
increased from 77.10\% at $L=0$ to 97.19\% at $L=5$.
With AP4+4 at the maximum terminal-selection limit, the
median prediction times were 0.0146~s for the proposed
method and 0.0808~s for Coop-BP.

\section*{Acknowledgment}
This research was supported in part by the JSPS KAKENHI 23H00474, 23H01409, the Telecommunications Advancement Foundation, and the Ministry of Land, Infrastructure, Transport, and Tourism (MLIT), Japan, through the Construction Technology Research and Development Subsidy Program (JPJ000094).

\bibliographystyle{IEEEtran}
\bibliography{references}

\end{document}